\documentclass[12pt]{article}
\input epsf.sty
\newcommand{\ra}{\rangle}
\newcommand{\la}{\langle}

\newcommand{\II}{{\cal I}}

\newcommand{\MM}{{\cal M}}
\newcommand{\BB}{{\cal B}}
\newcommand{\CC}{{\cal C}}
\newcommand{\OO}{{\cal O}}

\newcommand{\SSS} {{\cal S}}

\newcommand{\wh}{\widehat}
\newcommand{\wc}{\check}

\newcommand{\RR}{{\cal R}}
\newcommand{\NN}{{\cal N}}

\newcommand{\WW}{{\cal W}}

\newcommand{\be}{\begin{equation}}
\newcommand{\ee}{\end{equation}}
\newcommand{\ben}{\begin{eqnarray}\displaystyle}
\newcommand{\een}{\end{eqnarray}}
\newcommand{\refb}[1]{(\ref{#1})}
\newcommand{\p}{\partial}
\newcommand{\sectiono}[1]{\section{#1}\setcounter{equation}{0}}

\begin{document}

{}~
\hfill\vbox{\hbox{hep-th/0202151}\hbox{CTP-MIT-3232}
\hbox{CGPG-02/1-3}
\hbox{PUPT-2020}
}\break

\vskip .6cm

\centerline{\large \bf
Star Algebra Projectors  
}

\vspace*{4.0ex}

\centerline{\large \rm Davide Gaiotto$^a$, Leonardo Rastelli$^a$,
Ashoke Sen$^b$ and Barton Zwiebach$^c$}

\vspace*{4.0ex}

\centerline{\large \it ~$^a$Department of Physics }

\centerline{\large \it Princeton University, Princeton, NJ 08544,
USA}

\centerline{E-mail: dgaiotto@princeton.edu,
        rastelli@feynman.princeton.edu}

\vspace*{1.8ex}

\centerline{\large \it ~$^b$Harish-Chandra Research
Institute}

\centerline{\large \it  Chhatnag Road, Jhusi,
Allahabad 211019, INDIA}
\centerline {and}
\centerline{\large \it Department of Physics, Penn State University}

\centerline{\large \it University Park,
PA 16802, USA
}

\centerline{E-mail: asen@thwgs.cern.ch, sen@mri.ernet.in}

\vspace*{1.8ex}

\centerline{\large \it $^c$Center for Theoretical Physics}

\centerline{\large \it
Massachussetts Institute of Technology,}

\centerline{\large \it Cambridge,
MA 02139, USA}

\centerline{E-mail: zwiebach@mitlns.mit.edu}

\vspace*{4.0ex}

\centerline{\bf Abstract}
\medskip

Surface states are open string field configurations  
which arise from
Riemann surfaces with a boundary and form a subalgebra of the
star algebra.  We find that a general class  
of star algebra
projectors arise from surface states where 
the open string midpoint reaches the boundary of the surface.
The projector property of the state and the split nature of
its wave-functional arise because of a nontrivial feature of conformal maps
of nearly degenerate surfaces.
Moreover, all such projectors  
are invariant under constant and opposite translations of their
half-strings. We show that the half-string 
states associated to these 
projectors are themselves surface states.  In addition to the sliver, we
identify other interesting projectors. 
These include a  butterfly state,
which is the tensor product of half-string vacua, and
a nothing state, where the Riemann surface collapses.  
\vfill \eject

\baselineskip=16pt

\tableofcontents

\sectiono{Introduction and Summary} \label{s1}

The elucidation of tachyon dynamics and D-brane
instabilities in string theory \cite{conj} has led to renewed
investigations in open string field theory (OSFT). In particular,
the star-algebra of open string fields, an associative
multiplication introduced in \cite{OSFT}, is the key algebraic
structure in this theory. Recently, a formulation of open
string field theory based on the tachyon vacuum -- vacuum string field
theory (VSFT) -- was proposed in \cite{0012251,0111129} and investigated 
in \cite{0102112}--\cite{0201177}. 
In this theory,  D-brane solutions were seen to correspond to open string
fields that were projectors of the star algebra, {\it i.e.} elements  
which square  to
themselves, at least in the matter sector
\cite{0102112}. With the ghost structure of VSFT determined in
\cite{0111129,0108150,0201015}, D-brane solutions are indeed seen to
correspond
strictly to projectors in the star algebra whose underlying conformal field
theory (CFT) involves a twisting of the reparametrization
ghost CFT \cite{0111129}.  Since no direct derivation of VSFT from
conventional OSFT has yet been given,  one must still
consider VSFT somewhat conjectural.  In fact, the present version
of VSFT, where the kinetic operator is a ghost insertion
with infinite strength at the open string midpoint, appears to be
an extremely simple but somewhat singular limit that arises from a
reparametrization that maps the open string to its midpoint \cite{0111129}.
We nevertheless expect that the connection between projectors of the
star algebra and classical solutions of string field theory is a key
property that will persist in other formulations of vacuum string field
theory.\footnote{Since the BRST operator in OSFT is complicated
and mixes matter and ghost sectors nontrivially, the equations
of motion of this theory do not appear to have the structure of
projector equations.}

One is therefore led to investigate the existence of
projectors of the star algebra of open string fields.
For many years, the only string field known to multiply
to itself was the identity string field. In fact, there
were heuristic reasons to believe that projectors should
be scarce in the star algebra.\footnote{We learned of such
ideas from E. Witten.}  
One can understand the difficulty in constructing projectors within 
the large class of field configurations  that arise
from path integration over fixed Riemann surfaces whose boundary
consists of a  parametrized open
string and a piece with  open string boundary conditions. 
Such string fields,  called surface states, are easily
star multiplied.\footnote{We shall be using the standard correspondence 
between quantum states of the first quantized 
theory and classical field configurations in the second quantized theory 
to refer to string field configurations as "states". In the same spirit we 
use the term ``wave-functional" to refer to
the functional of the string coordinates 
that represents the classical string field configuration.}
One glues the right-half of 
the 
open string in the first surface to the left-half of the open string in
the second surface, and the surface state corresponding to the
glued surface is the desired product. It is clear from this
description that multiplication of a surface state to itself
leads to a surface state that looks different from the initial
state. This is the reason why it  seemed  difficult, in general,  to find 
projectors.

\medskip
Leaving aside the identity string field, 
the first projector of the star algebra to be found was the
sliver state \cite{0006240,0008252,0102112}. It circumvents the above
mentioned
difficulty in an interesting way. Consider wedge states,
surface states where the Riemann surface is an angular sector
of the unit disk, with the left-half and the right-half of the
open string being the two radial segments, and the unit radius arc
having the open string boundary conditions. A wedge state is
thus defined by the angle at the open string midpoint, and this
angle simply adds under star multiplication, as is readily
verified using the gluing prescription. The identity string field
is the wedge state of zero angle, and the sliver is the wedge
state of infinite angle! The addition of infinity to infinity is
still infinity, and the sliver does star multiply to
itself.\footnote{The only question here is whether the sliver defined
as the infinite angle limit of a sector state exists. It does, as
seen in \cite{0006240}, explained in detail in \cite{0105168} and
confirmed
numerically.}

\medskip
The sliver state was later recognized to be a string wave-functional
that is split: it is the product of a functional of the
left-half of the string times the same functional of the
right-half of the open string.  From this viewpoint, however,
it seemed surprising that projectors would be hard to find:
any symmetric split string wave-functional could serve as a projector.
At least algebraically other projectors could easily be
constructed by transforming the sliver by the action of
star conjugation. Geometrically, however, it was not clear 
if there are  
other projectors which
could be simply interpreted as other surface
states.

\begin{figure}[!ht]
\leavevmode
\begin{center}
\epsfysize=5cm
\epsfbox{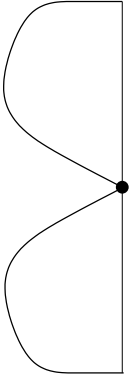}
\end{center}
\caption{The generic kind of surface state providing a projector
of the star algebra. The open string is the vertical boundary, 
and the open string midpoint is shown with a
heavy dot. The rest of the boundary has open string boundary 
condition. Note that this part of the boundary 
touches the open string midpoint.}
\label{0ff}
\end{figure}

\medskip
In this paper we find large classes of projectors that
indeed arise as surface states, and we explain the general
mechanism by which they all evade the heuristic argument
sketched earlier.  
All the projectors we find correspond
to surface states where in the Riemann surface,
the open string midpoint reaches the boundary where the
open string boundary conditions are imposed (this is also the
case for the sliver). 
The general situation is illustrated in  figure \ref{0ff}.
The vertical boundary is the
open string, and its midpoint is indicated by a heavy dot.
The rest of the boundary has open string boundary condition. 
More precisely, one
can regulate the surface state by letting the open string midpoint
reach the boundary when the regulator is removed.  We explain
that in the limit as the regulator is removed, the string wave-functional
splits into a product of functionals. In considering the
projector property, we examine the gluing of two regulated projector
surface states.
The gluing of the two regulated surfaces does not give a surface that
looks like the original one, but rather, a
surface that looks like the original one plus a short neck connecting
it to an additional disk. We will explain that, conformally speaking,
the  short neck and the extra disk are in fact negligible
perturbations in the sense that the resulting
surface is accurately conformally equivalent to the original one without
the extra disk. The agreement becomes exact as the regulator
is removed. We consider this a key insight in the present paper. It
shows how conformal equivalence is subtle enough to circumvent the
heuristic arguments against the existence of projectors.

\medskip
All the above are rank one projectors, and are expected to
be related by star conjugation.  Nevertheless there are special
projectors that deserve special attention, for they satisfy a
number of unusual conditions that may be of some relevance.
In addition to the  
sliver, our studies have uncovered two 
special projectors -- the butterfly and the 
`nothing' state.  We now summarize the
special properties of these  three projectors.

For the sliver $|\Xi\rangle$ we have:

\begin{itemize}

\item  It is the projector that arises by repeated star multiplication
of the SL(2,R) vacuum, namely  $|\Xi \rangle
= \lim_{n\to \infty} (|0\rangle)^n$.

\item  It is a projector whose Neumann matrix in the oscillator
representation commutes with those defining the star product.

\item It is the limit element of a sequence of surface states,
the wedge states, defining an abelian subalgebra of the star algebra.
The sliver state is annihilated by the star algebra derivation
$K_1 = L_1 + L_{-1}$.

\end{itemize}

The properties of the ``butterfly" state were 
announced in \cite{0111129}:  

\begin{itemize}

\item  It has an extremely simple presentation in the
Virasoro basis.  It is just $\exp(-{1\over 2} L_{-2}) |0\rangle$.
It is annihilated by the star derivation $K_2 = L_2 - L_{-2}$.

\item  Its wave-functional is  the product of {\it vacuum} wave-functionals
for the left-half and  the right-half of the string.
Thus, it is the simplest
projector from the viewpoint of half-strings.

\item  This is the state that appears to arise 
when considering the projector equations in the level expansion.

\end{itemize}

We constructed a family of projectors, all of them generalized
butterflies, that interpolate from the sliver to the above
canonical butterfly.  The family can be continued beyond this
canonical butterfly state up  
to a projector that we call the ``nothing state".
The nothing state has the following properties:  
\begin{itemize}
\item As we reach this state the Riemann surface becomes vanishingly
small.
\item 
It is annihilated by
all the derivations $K_{2n} = L_{2n} -
L_{-2n}$.

\item 
It has a constant wave-functional.

\end{itemize}

Our general discussion  shows that the condition that
the open string midpoint touches the boundary ensures  
that projectors have split wave-functionals, that is, wave-functionals
that factorize into a product of functionals each involving
a half-string.  We also show that the half-string states
associated to surface state projectors are themselves
surface states defined with the same boundary condition
as the original projector, 
and give an explicit algorithm for the 
construction of such 
half-string states (see \refb{toptohalf} and \refb{halftotop}). 
For example,  
our construction explains why the butterfly is the state
corresponding to the tensor product of half 
string vacua with 
Neumann boundary condition at both ends if the original butterfly is 
defined with Neumann boundary condition.
We believe this is
an interesting insight into half-string formalisms, where
boundary conditions at the string midpoint are subject to debate,
and little geometrical understanding is available. 
We also point out a 
subtletly. 
When we say that a surface
state is 
{\it defined}  using a boundary condition, this means that
the functional integral defining the state is done imposing
the boundary condition in question along the boundary of the
surface. A surface state $|\Sigma\rangle$
{\it defined} 
 using a given boundary condition, say of Neumann type, may not
necessarily satisfy  
this condition, namely, expectation values of the
operator  $ \partial_\sigma  X(\sigma)$  
on the state $|\Sigma\rangle$ may 
not necessarily vanish at
$\sigma = 0,\pi$.  
This may happen when the boundary of the
surface $\Sigma$ has a corner type singularity at the string endpoints. 
While such corner type singularities are not common in familiar
surface states, they are generic for half-string states.  
Thus, for example,
we find that unlike 
the canonical butterfly half-string state,
the nothing half-string state satisfies 
a Dirichlet boundary condition at the point corresponding to the full
string midpoint.

We also show that, just as for the sliver \cite{0111069}, 
all projectors 
defined with Neumann boundary condition 
have wave-functionals invariant under constant
and opposite translations of the half-strings. This implies
that the Neumann matrices associated to projectors all have
a common eigenvector, the $\kappa=0$ eigenvector of $K_1$.
This simply follows, as we explain in the text,
because the associated half-string surface states 
carry no momentum.

\medskip
This paper is organized as follows.  
We begin in section \ref{s2} by reviewing various geometrical
presentations of surface states, and various concrete algebraic
representations of the states.  
Section \ref{sdegenerate} is devoted to a discussion of
some properties of conformal maps and conformal field
theories which we use in later analysis. In particular we discuss
factorization properties of conformal field theory correlators on a
pinched disk that is about to split into two disconnected disks.
Although we do not
attempt to state precisely the way in which the surface must be pinched to
achieve the desired factorization property, we state a conjecture that we
expect to
hold, and would guarantee the key properties. We also
illustrate the conjecture with an example.

In section \ref{s3a} we explain in general terms our understanding of the
construction of  split
wave-functionals.
In particular we argue, based on the results of sections \ref{s2}, 
\ref{sdegenerate} that the
surface state associated with a  
conformal map that sends the
string midpoint to a point on the boundary of the disk will give a state
whose wave-functional factorizes into a functional of the left
half of the string and a functional of the right half of the string. The
analysis in this section is done in the context of a general boundary
conformal field theory. We show how to construct the half-string 
{\it surface } states
corresponding to split wave-functionals.

In section \ref{new5} we explain in general terms that the same
condition that gives rise to split wave-functionals also guarantees that
the corresponding states are projectors.
We show that the projectors are of rank one.
In addition, we prove that all surface state projectors have Neumann matrices
that share a common eigenvector.

In section \ref{s2b} we describe in detail the butterfly state.
We show how to regulate it and prove explicitly that the
state satisfies the projector equation by constructing
the requisite conformal maps. We also use the method of section \ref{s3a}
to show the factorization property of the butterfly wave-functional,
and explicitly determine the functional of the left and the
right-half of the string to which the butterfly wave-functional
factorizes. Both of these turn out to be the wave-functional of the vacuum
state. Section \ref{snothing} is devoted to a similar study of 
the nothing state.

In section \ref{s2c}
we introduce the family
of generalized butterflies parametrized by a parameter $\alpha$,
interpolating from the sliver at $\alpha=0$
up to the `nothing' state at $\alpha=2$, passing through the butterfly at
$\alpha=1$. We show explicitly that for every $\alpha$ the associated
surface state is a projector. The nothing 
state at $\alpha=2$ is particularly
interesting, since after removing the local coordinate patch the
corresponding surface has vanishing area. 
We also show
explicitly that the wave-functionals of these generalized butterfly states
factorize into functionals of the left- and the right-half string
coordinates, and determine these functionals.

In section \ref{s5} we discuss   
additional simple projectors, which just as the butterfly, can
be represented by the exponentiation of a single Virasoro operator.
We discuss some properties of these projectors, and sketch the 
construction of certain star-subalgebras.
In section \ref{s4} we discuss butterfly states associated to
general boundary conformal field theories (BCFT's), 
represented as a state 
in the state space of some fixed
reference BCFT. This generalizes previous arguments that
were known to hold for the sliver state.  Finally we offer some
concluding remarks in section \ref{concl}. 
We include 
an appendix 
where we give the explicit numerical computation
of the $*$-product of the butterfly state with itself to show that it
indeed behaves as a projector of the $*$-algebra. We also test
sucessfully the expected rank one property of the projector. 

\medskip
Related but independent research on the matter of projectors and 
butterfly states has been published recently by
M.~Schnabl \cite{schnablNEW}.

\sectiono{Surface States -- Presentations and Representations} 
\label{s2}

In this section we shall discuss general properties of surface states. 
After reviewing the geometric description of surface states in various 
coordinates, we discuss the  
computation of star products and inner products of 
surface states in the geometric language. We then 
discuss  the explicit
operator, oscillator and functional representations
 of general surface states.

\subsection{Reviewing various presentations} \label{s2.1}

For the purposes of the arguments in this paper we will
review the various coordinate systems used to describe
surface states. A surface state $\langle \Sigma|$ 
for the present
purposes arises from a Riemann surface $\Sigma$ with the topology of a
disk, with a marked point $P$, the puncture, lying on the boundary
of the disk, and a local coordinate around it.

\medskip
\noindent
{\it The $\xi$ coordinate.}  This is the local coordinate.
The local coordinate, 
technically speaking is a map from the canonical half-disk
$|\xi| \leq 1, \Im (\xi) \geq 0$
into the Riemann surface $\Sigma$, where the boundary $\Im (\xi) = 0,
|\xi | < 1$ is mapped to the boundary of $\Sigma$
and $\xi=0$ is mapped to the puncture $P$.  
The open
string is the $|\xi| = 1$ arc in the  half-disk.
The point $\xi=i$ is the string midpoint. The surface $\Sigma$
minus the image of the canonical $\xi$ half-disk will be 
called $\RR$.
Using any {\it global coordinate} $u$ on the disk representing $\Sigma$,
and writing 
\be u = s(\xi)\,, \quad \hbox{with} \quad s(0)= u(P)\,,
\ee
the surface  
state
$\la\Sigma|$ is then defined through the relation:
\be \label{edefsurn}
\la\Sigma|\phi\ra = \la  s\circ \phi(0)\ra_{\Sigma}\, ,  
\ee
for any state $|\phi\ra$. Here
$\phi(x)$ is the vertex operator corresponding to the state
$|\phi\ra$ and $\la~\ra_\Sigma$ denotes the correlation function on the
disk $\Sigma$. There is nothing special about a specific choice of
global coordinate $u$, and the state $\langle \Sigma|$ built with
the above prescription does not change under a conformal map taking
$u$ to some other coordinate and $\Sigma$ into a different looking
(but conformally equivalent) disk. Nevertheless there are particularly
convenient choices which 
we now discuss in detail.

\medskip\noindent
{\it The $z$- presentation.}  In this presentation the Riemann
surface $\Sigma$ is mapped to the {\it full} upper half $z$-plane, with
the puncture lying at $z=0$. The image of the canonical $\xi$
half-disk  is some region around $z=0$. Thus if $z=f(\xi)$, we have
\be \label{edefsurcan}
\la\Sigma|\phi\ra = \la f\circ \phi(0)\ra_{UHP}\, .
\ee

\medskip\noindent {\it The} $\wh z$-{\it presentation.} In this
presentation the Riemann surface $\Sigma$ is mapped such that the image of
the canonical $\xi$ half-disk is the full strip $|\Re (\wh z)|\leq
{\pi/4}, \Im (\wh z ) \geq 0$, with $\xi=i$ 
mapping to $\wh z =i\infty$,
and the open string mapping to the vertical 
half lines at $\Re(\wh z) = \pm {\pi\over 4}$. 
This is
implemented by the map
\be \label{xx22}
\wh z = \tan^{-1} \xi\,.
\ee
The
rest $\RR$ of $\Sigma$ will take some definite shape that will typically
fail to coincide with the full upper-half $\wh z$ plane. This shape
actually carries the information about the surface $\Sigma$.

\medskip\noindent
{\it The $\wh w$- presentation.}  In this presentation the Riemann
surface $\Sigma$ is mapped such that the image of the canonical $\xi$
half-disk is the canonical half disk $|\wh w|\leq 1,
\Re (\wh w) \geq 0$, with $\xi=0$ mapping to $\wh w =1$. This
is implemented by the map
\be \label{x2}
\wh w = {1+i\xi\over 1 - i \xi} \equiv h(\xi)\, .
\ee
The rest
$\RR$ of the surface will take some definite shape $\wh\RR$ 
in this
presentation. This shape actually carries the information about the
surface $\Sigma$. We also note that the $\wh w$-presentation
and the $\wh z$ presentation are related as
\be
\label{hwhz}
\wh w = \exp ( 2 i \wh z)\,.
\ee

\medskip\noindent
{\it The $\xi$- presentation.}  
In this
presentation the Riemann surface $\Sigma$ is mapped into the $\xi$-plane
by extending to the whole surface $\Sigma$ the map that 
takes the neighborhood of
the puncture $P\in \Sigma$ into the $\xi$-half-disk.  This extended map, of
course, may require branch cuts. In this presentation 
the surface is the canonical
$\xi$ half-disk plus 
some region in the $\xi$-plane whose 
shape carries the information of the state.
We call $\Sigma_\xi$ the surface in 
this presentation. In this case the equation
defining the state takes a particularly 
simple form since no conformal map
is necessary
\be
\la\Sigma|\phi\ra = \la  \phi(0)\ra_{\Sigma_\xi}\,.
\ee
The $\xi $ presentation can be obtained from
the $\wh w$ presentation by the action of $h^{-1}$.

\subsection{ Inner products and star-products of surface states}  
\label{sinner}

\begin{figure}[!ht]
\leavevmode
\begin{center}
\epsfysize=5cm
\epsfbox{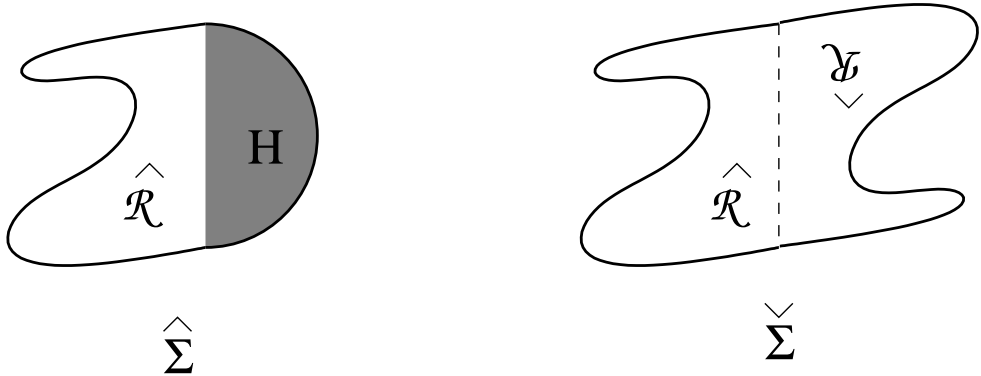}
\end{center}
\caption{The geometry involved in computing the inner product of a
surface state $|\Sigma\ra$ with itself.
} \label{fx1} \end{figure}

In this subsection we will address 
the 
computation of correlation functions of the 
form $\la \Sigma | \prod_{i=1}^n \OO_i(\xi_i) |\Sigma\ra$
with $\xi_i$'s lying on the unit circle.  
These are  
inner products of surface states with operator insertions.
Such computations will play a role in our later analysis 
of split wave-functionals and half-string states. We will also 
discuss star products
of surface states.

These computations are 
particularly simple in the representation of the surface $\Sigma$ 
in the  $\wh w$  coordinate system. 
Let $\wh \RR$ and $\wh
\Sigma$ denote the
images of $\RR$ and $\Sigma$ in the $\wh w$ plane.  Then we can rewrite
eq.\refb{edefsurn} as
\be \label{yx1}
\la\Sigma| \phi\ra = \la h\circ\phi(0)\ra_{\wh\Sigma}\, ,
\ee
where $h$ has been defined in
eq.\refb{x2}. To compute $\la \Sigma | \prod_{i=1}^n \OO_i(\xi_i) 
|\Sigma\ra$
we begin with two
copies of $\wh\Sigma$, remove the local coordinate patches from each so
that we are left with two copies of $\wh\RR$, and then simply construct a
new
disk by gluing the left-half string of the first disk to the right
half-string of the second string and vice versa, as shown in
Fig.\ref{fx1}.
If we denote the new disk by $\wc\Sigma$
then we have
\be \label{xx2}
\la \Sigma|  \prod_{i=1}^n \OO_i(\xi_i) |\Sigma\ra =
\la\prod_{i=1}^n h\circ \OO(\xi_i)\ra_{\wc \Sigma}\, .
\ee
The $h\circ\OO_i(\xi_i)$
factors are inserted at the images
of
the points $\xi_i=e^{i\sigma_i}$ in the $\wh w$ plane, {\it i.e.} on the
imaginary axis.

\begin{figure}[!ht]
\leavevmode
\begin{center}
\epsfysize=5cm
\epsfbox{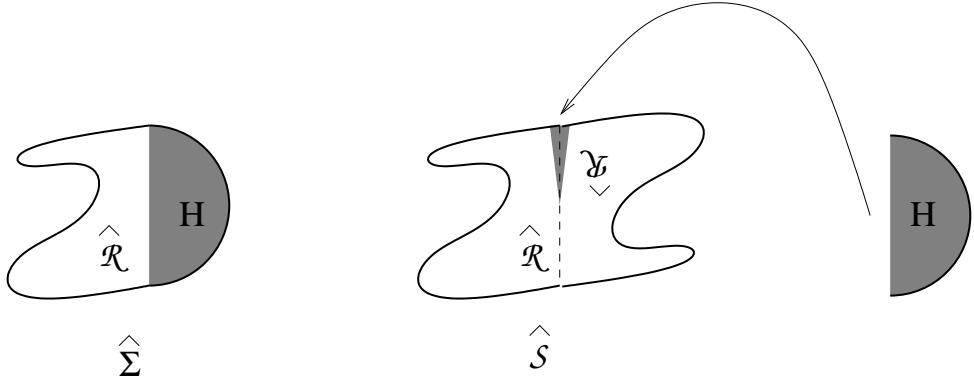}
\end{center}
\caption{The geometry involved in computing the star product of a
surface state with itself. The local coordinate patch, shown as the shaded
half-disk to the right, is to be glued to the shaded region of the diagram
representing $\wh\SSS$.} \label{fx3}
\end{figure}

We shall also need to compute the star product of a surface state with 
itself. This is again simple in the $\wh w$ coordinate system.
We begin with
two copies of the disk
$\wh\Sigma$, remove the local
coordinate patch from the second $\wh\Sigma$, and
then glue the right half-string of the first $\wh\Sigma$ with the left
half-string of the second $\wh\Sigma$. The result is a new disk $\wh\SSS$,
as shown in Fig.\ref{fx3}. As indicated in the figure, the local 
coordinate
patch is glued in and thought as part of $\wh\SSS$.
The
surface state $|\SSS\ra$ associated with the new surface $\wh\SSS$ gives
$|\Sigma*\Sigma\ra$.
Thus we have:
\be \label{estar}
\la \Sigma * \Sigma|\phi\ra = \la h\circ \phi(0)\ra_{\wh \SSS}\, .
\ee

\subsection{Operator representation of surface states} \label{soper}

We shall now review the 
explicit representation of surface states in terms of 
Virasoro 
operators acting on the SL(2,R) invariant vacuum, rather than 
the implicit representation through correlators given eq.\refb{edefsurcan}.
Using the SL(2,R) invariance of the upper half plane, we can make the map 
$f$ appearing in eq.\refb{edefsurcan} satisfy $f(0)=f''(0)=0$, 
$f'(0)=1$.
We can write the corresponding surface state $\la\Sigma|$ 
as
\be \label{Uf}
\la \Sigma | = \la 0| U_f \equiv \la 0| \exp\left(\sum_{n=2}^\infty
v_n^{(f)} L_n \right) \, ,
\ee
where the coefficients $v_n^{(f)}$ are determined
by the condition that the vector field 
\be \label{edefvxi}
v(\xi) = \sum_{n=2}^\infty v_n^{(f)} \xi^{n+1}\, ,
\ee
exponentiates
to $f$,
\be \label{vdef}
\exp \left( v(\xi) \p_\xi \right) \xi
= f(\xi) \,.
\ee
We now
consider the one-parameter family of maps
\be \label{fbeta}
f_{\beta} (\xi) =  \exp \Bigl(\beta \, v(\xi) {\partial\over \partial \xi}
\Bigr)\, \xi
    \,.
\ee
This definition immediately gives
\be \label{vf}
 \frac{d}{d \beta}  f_\beta(\xi) = v(f_\beta (\xi)) \,.
\ee
Solution to this equation, subject to the boundary condition
$f_{\beta=0}(\xi)=\xi$, gives:
\be \label{egenc1b}
f_\beta(\xi) = g^{-1}(\beta + g(\xi))\, ,
\ee
where
\be \label{egenc2}
g'(\xi) = {1\over v(\xi)}\, .
\ee
Thus
\be \label{egenc1}
f(\xi) = g^{-1}(1 + g(\xi))\, .
\ee
Equations \refb{egenc2} and \refb{egenc1} readily give $f(\xi)$ if $v(\xi)$
is known. Alternatively, they also determine $v(\xi)$ in terms of $f(\xi)$,
although not explicitly, since eqn.~\refb{egenc1} is in general hard to
solve for $g$. When a solution for $v(\xi)$ is available,
eqn.\refb{Uf}  gives the operator expression for $|\Sigma\ra$.

\subsection{Oscillator representation of surface states} 

If the BCFT under consideration is that of free scalar fields  
with Neumann boundary condition 
describing D-25-branes 
in flat space-time, we can also represent 
the state in terms of the oscillators associated with the scalar fields.
For simplicity let us restrict our attention to the matter part of the 
state only.
If $a_m$, $a_m^\dagger$ denote the annihilation and creation operators 
associated with the scalar fields, then we have:
\be \label{soscill}
|\Sigma\ra =\exp \left(  -\frac{1}{2} \sum_{m,n=1}^\infty a^\dagger_m
V^f_{mn} a^\dagger_n \right) |0 \ra \,.
\ee
where we have suppressed spacetime indices and \cite{lpp}
\be  \label{evmn}
V^f_{mn} = {(-1)^{m+n+1}\over \sqrt{mn}}
\oint_0{dw\over 2\pi i}
\oint_0{dz\over 2\pi i}
\,{1\over z^m w^n} {f'(z)
f' (w)\over (f(z) - f(w))^2}\,.
\ee
Both $w$ and $z$ integration contours are circles around the origin,
with the $w$ contour lying outside the $z$ contour, and both contours
lying inside the unit circle. 
In eq.\refb{soscill} we have also omitted an overall normalization factor.

\medskip  
We now show that when the vector field $v(\xi)$ generating the
conformal map $f(\xi)$ is known (see the discussion in the previous
subsection) the above integral expression for the matrix
$V^f$ of Neumann coefficients
can be given an alternate form which is sometimes easier to evaluate.
For this purpose,  we now consider the matrix $V (\beta)$
associated to the family of maps \refb{fbeta},
and rewrite \refb{evmn} as
\be
V_{mn}(\beta) \equiv V_{mn}^{f_\beta} =  {(-1)^{m+n+1}\over \sqrt{mn}}
\oint_0{dw\over 2\pi i}
\oint_0{dz\over 2\pi i}
\,{1\over z^m w^n} \frac{\partial}{\partial z} \frac{\partial}{\partial w}
\log(f_\beta(z) - f_\beta(w))\, .
\ee
Taking a derivative with respect to the parameter
$\beta$,
\ben \label{Vbetalog}
\frac{d}{d \beta}  V_{mn}(\beta) & = & \frac{(-1)^{m+n+1}}{  \sqrt{mn}}
\oint_0{dw\over 2\pi i}
\oint_0{dz\over 2\pi i}
\,{1\over z^m w^n} \frac{\partial}{\partial z} \frac{\partial}{\partial w}
\frac{\partial }{\partial \beta}
\log (f_\beta(z) - f_\beta(w))\, \\
& = &  \frac{(-1)^{m+n+1}}{  \sqrt{mn}}
\oint_0{dw\over 2\pi i}
\oint_0{dz\over 2\pi i}
\,{1\over z^m w^n} \frac{\partial}{\partial z} \frac{\partial}{\partial w}
\left( \frac{  v(f_\beta(z)) - v(f_{\beta}(w))}{f_\beta(z) -
f_\beta(w)} \right)\, ,
\nonumber
\een
where we have exchanged the order of derivatives and used \refb{vf}.
Integration by parts in $z$ and $w$ then gives
\be \label{Vbeta}
\frac{d}{d \beta}  V_{mn}(\beta) = {(-1)^{m+n+1}  \sqrt{mn}}
\oint_0{dw\over 2\pi i}
\oint_0{dz\over 2\pi i}
\,{1\over z^{m+1} w^{n+1}}  \frac{  v(f_\beta(z))
- v(f_{\beta}(w))}{f_\beta(z) -
f_\beta(w)} \, .
\ee
This is a 
general formula that we  have 
found  
useful in concrete computations.
If the above matrix is calculable,
the desired Neumann coefficients $V_{mn}(\beta=1)$
are then readily obtained
by integration over $\beta$.

\subsection{Wave-functionals for surface states} \label{s3.1} 

In this subsection
we wish to establish a dictionary
between the geometric interpretation of
surface states as one-punctured disks
and their representation as
wave-functionals of open string
configurations. For this we consider open strings on a D-25-brane in flat 
space-time so that we have Neumann boundary condition on all the fields, 
and write, for a surface state associated to the map $z= f(\xi)$ 
\be \label{ewave1}
\la \Sigma | X(\sigma) \ra = {\cal N}_f \exp \left( -\frac{1}{2}
\int_0^\pi \int_{0}^\pi  d\sigma d \sigma'
X(\sigma) A_f(\sigma, \sigma') X(\sigma') \right) \, .
\ee
Here we have suppressed the Lorentz indices and used the fact that the
wave-functional is in fact gaussian. This follows since
the vacuum $| 0 \rangle$ is represented by a gaussian
wave-functional, and the action of $U_f$ preserves this property. The
normalization constant $\NN_f$ is chosen so that $\la\Sigma|\Sigma\ra=1$.
The wave-functional can also be represented in terms of modes:
\be \label{ewave}
\la \Sigma | X \ra = {\cal N}_f \exp( -\frac{1}{2} \sum_{n,m=1}^\infty
X_n A^{(f)}_{nm} X_m )\,,
\ee
where we have adopted the convention of ref.\cite{gross-jevicki} to define
the modes $X_n$:
\be \label{eexpan}
X(\sigma) = X_0+\sqrt{2} \sum_{n=1}^\infty X_n \cos ( n \sigma) \,.
\ee
For simplicity we are
considering the case where the coordinate $X$ has Neumann boundary
condition. Using eqs.\refb{ewave1}, \refb{ewave}, \refb{eexpan} we see
that
$A_f(\sigma,\sigma')$ is related to $A^{(f)}_{nm}$ through the
relation:
\be \label{modecont}
A_f(\sigma,\sigma')={2\over \pi^2} \sum_{m,n\ge 1} A^{(f)}_{nm}
\cos(n\sigma)
\cos(m\sigma')\, .
\ee
Note that $X_0$ does not appear in the expression for the  
wave-functional in \refb{ewave}, since all surface states are
translationally invariant.
We shall from now on restrict to twist-even surface states,
which is equivalent to the condition that $f$ is an odd
function.

The relation between the oscillator representation \refb{soscill} and the 
wave-functional representation \refb{ewave} is given by the following 
relation 
between the matrices $V^f$ and $A^{(f)}$\cite{0105058,0105059}: 
\be \label{evfaf}
A^{(f)} = 2 E^{-1}\,{1-V^f\over 1+ V^f}\, E^{-1} \,, 
\qquad E_{nm} = 
\delta_{mn} \sqrt{2\over n}\, .
\ee

We want to determine $A_f(\sigma, \sigma')$ from $f$.
To this end, we evaluate the normalized correlator
\be \label{Bdef}
B_f(\sigma_1, \sigma_2) \equiv
\la \Sigma | \partial_{\sigma_1}\hat X(\sigma_1) \partial_{\sigma_2}\hat
X(\sigma_2) |\Sigma \rangle
\ee
in two different ways. First, we use the wave-functional
representation and obtain 
\ben \label{Bgauss}
B_f (\sigma_1, \sigma_2 ) & =&
\NN_f^2\int {\cal D} X
\exp\left(-
   \int_{0}^\pi  d\sigma d \sigma'
X(\sigma) A_f(\sigma, \sigma') X(\sigma') \right)
\partial_{\sigma_1} X(\sigma_1) \partial_{\sigma_2} X(\sigma_2)  \nonumber
\\
& =& \frac{1}{2 } \partial_{\sigma_1} \partial_{\sigma_2} (A_f^{-1}
(\sigma_1, \sigma_2))\,.
\een
Here the inverse kernel $A_f^{-1}(\sigma_1, \sigma_2)$
is defined by
\be \label{eker}
\int_0^\pi d \sigma  A_f( \sigma_1, \sigma) A_f^{-1}(\sigma, \sigma_2) =
\delta(\sigma_1 - \sigma_2) - {1\over \pi}\,.  
\ee
The constant $1/\pi$ in the above equation
represents the contribution from the zero mode
part.
Since $A_f$ does not depend on the zero modes, it has an inverse only in
the subspace spanned by the functions $\cos(n\sigma)$ for $n\ne 0$.
Thus $(A_f^{-1})_{mn}$ 
is nonzero only for $m,n\geq 1$. 

\medskip
In the second computation,  we  interpret  
(\ref{Bdef}) as a  CFT correlator on an appropriate
Riemann surface following the procedure of section \ref{sinner}.
This is nicely  done using the $\wh z$ presentation (see section \ref{s2.1}) 
where the surface will occupy  a region $\CC_f$ in
the $\wh z$ plane, and 
the local coordinate patch 
is the strip $\Im(\wh
z)\ge 0$, $-\pi/4\le \Re(\wh z) \le \pi/4$. As usual the vertical line
corresponding to
$\Re(\wh z)=\pi/4$ is the image of the left-half of the string, and the
vertical line corresponding to $\Re(\wh z)=-\pi/4$ is the image of the
right-half of the string.
In order to compute a correlation function of the form $\langle
\Sigma|\cdots|\Sigma\rangle$, we simple start with
two copies of $\CC_f$,  strip off the local coordinate patch from each of
them, and compute
the correlation function on the surface obtained by gluing the left
half-string on the first $\CC_f$ with the right half-string on the second
$\CC_f$ and vice versa. Thus the right hand side of \refb{Bdef}
can be computed by this
method. If we denote the new surface by $\wc C_f$,
then $B_f$ is given by  
\be \label{expli}
B_f (\sigma_1,\sigma_2) = \p_{\sigma_1} \p_{\sigma_2}
\la  X(\wh z_1) X(\wh z_2)
\ra_{\wc C_f}\, ,
\ee
where $\wh z_i$ are the images of the points $\xi = e^{i\sigma_i}$. 
Comparing \refb{Bgauss} with \refb{expli} we can determine
$A_f^{-1}(\sigma_1, \sigma_2)$ and hence the wave-functional of the
surface state.

Let us illustrate this for
the vacuum state $| 0 \rangle$. In this case
the CFT computation is immediate, since
we can directly compute the correlator
(\ref{Bdef}) using the OPE's of X's on the UHP,
\ben \label{Bvac}
B_{|0 \rangle} (\sigma_1, \sigma_2 ) & =&  \partial_{\sigma_1}
\partial_{\sigma_2} \Big(-\frac{1}{2}
\log( |e^{i \sigma_1} - e^{i \sigma_2}|^2
| e^{i \sigma_1} - e^{-i \sigma_2}|^2) \Big) \nonumber \\
&=&
- \partial_{\sigma_1}
\partial_{\sigma_2} \log( 2 \cos(\sigma_1) - 2 \cos(\sigma_2))\nonumber
\\
&=& \partial_{\sigma_1}
\partial_{\sigma_2} \sum_{n=1}^\infty \frac{2}{n} \cos (n \sigma_1)
\cos( n \sigma_2) \,.
\een
Using (\ref{Bgauss}) and inverting the kernel, we get
\be
A_{| 0 \rangle} (\sigma_1, \sigma_2)
= {2\over \pi^2} \sum_{n=1}^\infty n \cos(n\sigma_1) \cos(n\sigma_2)\,.
\ee
This gives  
\be
A^{| 0 \rangle} _{nm} = n \delta_{nm} \,.
\ee
This is the expected result, as can be seen by using eq.\refb{evfaf} with 
$V^f=0$.

\sectiono{Conformal Field Theories on Degenerate Disks} 
\label{sdegenerate}

In this section we shall consider 
some properties of degenerate disks 
involving singular conformal maps, and of conformal field theories on such 
degenerate disks.  

\subsection{A conformal mapping claim} \label{s2.2}

\begin{figure}[!ht]
\leavevmode
\begin{center}
\epsfysize=3.5cm
\epsfbox{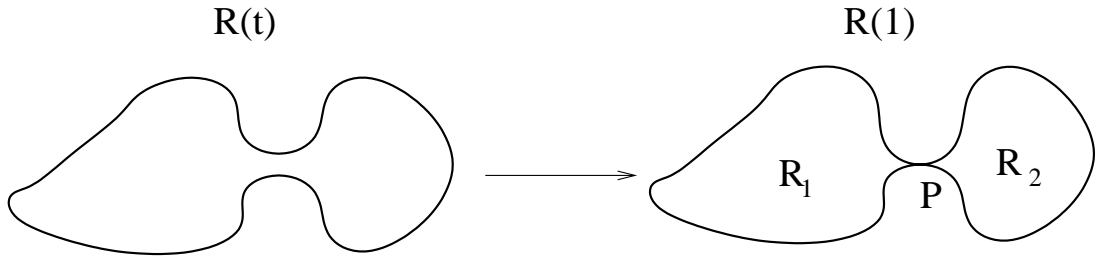}
\end{center}
\caption{The surface $R(t)$ is pinching for $t=1$. The pinching
point $P$ separates the regions $R_1$ and $R_2$ of the
surface $R(1)$.}
\label{02ff}
\end{figure}

Consider a set of surfaces parametrized by $t\in [0,1]$.
For each $t$
different from one, the surface  
$R(t)$ is a finite region of the complex plane with the
topology of a disk (see figure \ref{02ff}). 
As $t$ goes to one the region 
varies smoothly throughout but  develops a
thin neck and at $t=1$ it pinches, breaking into two pieces  
$R_1$ and $R_2$, both of which are finite disks.
Let $P$ denote the pinching point, common
to $R_1$ and $R_2$, and assume there are no other pinching
points. Because disks can always be mapped to disks, any $R(t)$ with
$t<1$ 
can be mapped to $R_1$. The map in fact is not unique
due to SL(2,R) invariance of the disk.  
We now claim that:

\medskip
\noindent
{\it Claim:
There exists a family of
conformal maps $m(t): R(t) \to R_1$, for $t\in [0,1]$,
continuous in $t$, where  $m(1)$ is the identity map over
$R_1$ and maps all of $R_2$ to $P$.}

\medskip

The intuition here is that as far as one of the sides
of the pinching surface is concerned, call it side one,
all that is going on on the other side, side two, can be viewed as
happening near the pinching point. The complete side two, lying on the
other side of the neck, can be mapped to a vanishingly small region
while the conformal map is accurately close to the identity on side
one.
An explicit example of this will be given next.

\subsection{A prototype example}

\begin{figure}[!ht]
\leavevmode
\begin{center}
\epsfysize=7cm  
\epsfbox{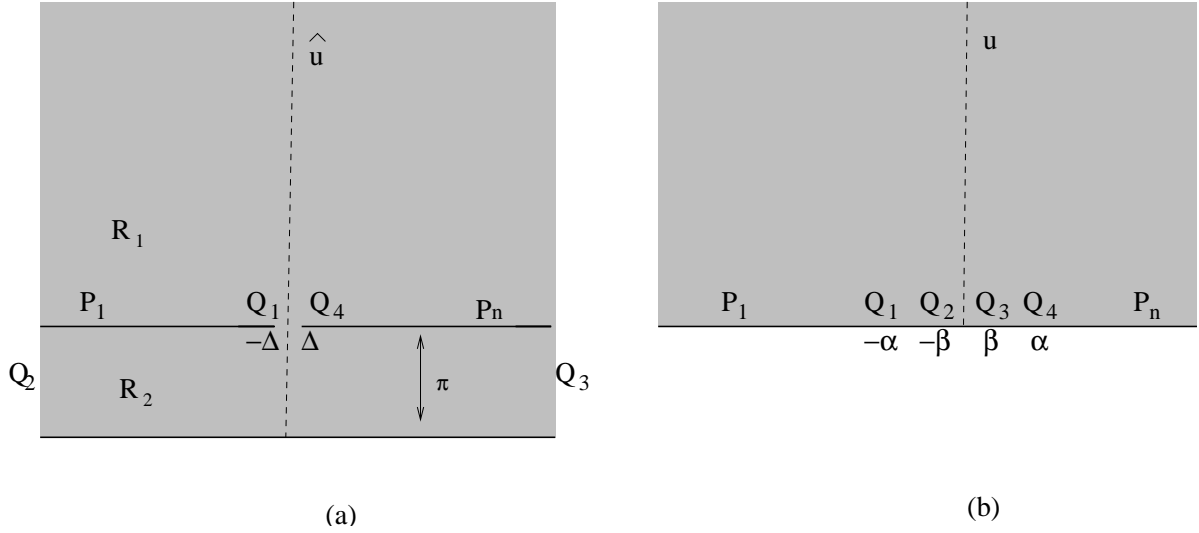}  
\end{center}
\caption{Illustration of a conformal map from the upper-half plane plus a
strip of width $\pi$ connected by a narrow neck (part (a)) to the upper-half
plane (part (b)).}
\label{ff1}
\end{figure}

To illustrate the claim in the above subsection
we consider the following situation.  Let a surface
$\Sigma$ with the topology of a disk be the
region in the $\hat u$-plane defined by 
$\Im (\hat u)\geq -\pi$  
with two cuts, both along the real axis, 
the first for real $\hat u \in (-\infty, -\Delta]$
and the second for real $\hat u \in [\Delta, +\infty)$ where $\Delta$
is a small positive real number. As illustrated as a shaded region
in figure \ref{ff1}(a), this region is the upper half plane, joined
through the
small interval $\hat u \in [-\Delta, \Delta]$
to an infinite horizontal
strip of width
$\pi$.  We have
also marked some special points $P_1, \cdots P_n$ at some
real coordinates $\hat u(P_i)$ with  $|\hat u (P_i)| \gg \Delta$, for
all $i$.

This surface, in the limit
$\Delta\to 0$, is pinching, and in the terminology of the claim,  the region
$R_1$ is the upper half
$\hat u$-plane, and the region $R_2$ is the horizontal infinite
strip of width $\pi$ in the lower half plane.  We will now 
show that the surface
can be mapped completely to the upper-half $u$ plane (figure \ref{ff1}(b))
such that the conditions in the claim are satisfied. Indeed, in the limit
$\Delta\to 0$, the map will become the identity over
the upper half plane ($R_1$) and will map all of
the lower strip ($R_2$) into a point. We will also see that
this is the limit of a family of maps that near
degeneration leave $R_1$ mostly unchanged (in a quantifiable way).

\medskip
The conformal map differential equation is readily written,
as it is of Schwarz-Christoffel type. Noting that 
the turning points $Q_1$ and  
$Q_4$ are points of turning angle $(-\pi)$, while $Q_2$ and $Q_3$ are
turning points of turning angle $(+\pi)$ we write:
\be
\label{theexamplemap}
d\hat u =  {u^2 -\alpha^2\over u^2 - \beta^2}  \,\,du\,.
\ee
The
turning points $Q_1, Q_2, Q_3, Q_4$ are mapped to the points
$\{-\alpha,-\beta, \beta, \alpha\}$ on the $u$-plane real axis, as shown
in \ref{ff1}(b).  
We have two parameters $\{\alpha, \beta\}$ and two conditions, one
defining the width of the strip, and the other specifying the separation
$2\Delta$ between $Q_1$ and $Q_4$ in the $\hat u$ plane. The normalization
above
was fixed so that for large $u \gg \{ \alpha, \beta\}$, we have
$d\hat u \sim du$, --   
a necessary condition for the map to become the identity
when
$u$ and $\hat u$ are large. The condition that the strip corresponding 
to $R_2$ has 
width $\pi$ demands that the residues of the above right hand side
at $u=\pm\beta$ be equal to $(\mp 1)$.  This gives
\be
\label{toyrc}
{\alpha^2 - \beta^2\over 2\beta } = 1 \,.
\ee
With this condition, the differential relation in \refb{theexamplemap}
becomes:
\be
\label{txamplemap}
d\hat u =  \Bigl( 1 - {1\over u-\beta} + {1\over u+\beta}\Bigr)
\,\,du\,.
\ee
By symmetry, we require that $u=0$ correspond to $\hat u = -i\pi$ and
thus we have that
\be
\Delta = P \int_0^\alpha \Bigl( 1 - {1\over u-\beta} +
{1\over u+\beta}\Bigr)\,
\,\,du\,,
\ee
where $P$ denotes principal value, which must be taken at $u=\beta$.
Evaluation gives
\be
\label{delte}
\Delta = \alpha + \ln \Bigl( {\alpha+ \beta\over \alpha-\beta}\Bigr) \,.
\ee
It is clear from this equation that to have $\Delta$ small we need
$\alpha$ small and $\beta \ll \alpha$. This, and  the constraint in
\refb{toyrc} can be satisfied with
\be
\alpha=\sqrt{2\beta+\beta^2}, \qquad \alpha,\beta\ll 1 \quad \to \quad
\beta\simeq
{1\over 2} \alpha^2\, .
\ee
Note that given these relations, \refb{delte} gives us
\be
\Delta = 2\alpha + {\cal O} (\alpha^2)\,, \quad \to \quad
\Delta \simeq  2\alpha\,.
\ee
This shows that the whole boundary of the strip $R_2$, which is
mapped to $u\in [-\alpha, \alpha]$, is indeed mapped to a vanishingly
small segment as $\Delta\to 0$.

 Finally, we confirm that the map goes to the identity map for $\{ u, \hat u
\}
\gg\Delta$ when $\Delta\to 0$.  For this purpose
integrating from
$\alpha$ to
$u>\alpha$ we have
\be
\hat u = \Delta + \int_\alpha^u \Bigl( 1 - {1\over u-\beta} +
{1\over u+\beta}\Bigr)\,
\,\,du\,,
\ee
which using \refb{delte} gives
\be
\hat u  = u + \ln \Bigl( {u+ \beta\over u-\beta}\Bigr) \,, \quad u>\alpha\,.
\ee
Since $u> \alpha \gg\beta$  we have
\be \hat u \simeq u + {2\beta\over u}
\simeq u + {\Delta^2\over 4u} \,\quad \to \quad u \simeq \hat u -
{\Delta^2\over 4\hat u} \,.
\ee
This confirms that away from the pinching area the map goes
to the identity map as $\Delta \to 0$.

\subsection{Conformal field theory and factorization} \label{s2.3}

Let us now consider a unitary  s
boundary conformal field theory on a surface $R(t)$
of the type described in section \ref{s2.2}
and consider a correlation function of the
form:
\be \label{econs1}
\la \prod_{i=1}^n \OO_i (z_i)\ra_{R(t)}\, .
\ee
$\OO_i$ could be bulk or boundary operators of
the theory.
We shall assume that all operators in the theory have dimension $>0$
except
the identity operator which has dimension zero. Let us now consider the
case where in the $t\to 1$ limit the points $z_1,\ldots z_m$ 
lie inside 
the disk $R_1$ and the points $z_{m+1},\ldots z_n$ lie inside the disk
$R_2$. In this limit, using the results of the
previous subsection we can map the disk $R$ to the disk $R_1$ in such a
way that the map is the identity inside $R_1$ and maps the 
whole of $R_2$ to a
point $P$. Thus the insertion points $z_{m+1},\ldots
z_n$ approach the point $P$. The correlation function on such a disk can
be evaluated
by picking up the leading terms in the operator product expansion of the
appropriate conformal transforms of the operators $\OO_{m+1},\ldots
\OO_n$. Since the lowest dimension operator in the theory is the identity
operator, we get:
\be \label{econs2}
\la \prod_{i=1}^n \OO_i (z_i)\ra_{R(t=1)}= \la \prod_{i=1}^m \OO_i
(z_i)\ra_{R_1} g(z_{m+1}, \ldots z_n)\, ,
\ee
where $g(z_{m+1}, \ldots z_n)$ is the function that appears in evaluating
the coefficient of the identity operator in the operator product expansion
of the appropriate conformal transforms of $\OO_{m+1},\ldots \OO_n$.

On the other hand, we could also carry out the analysis using a different
conformal transformation that maps the disk $R(1)$ to $R_2$ and maps the
disk $R_1$ to the point $P$. In this case we have
\be \label{econs3}
\la \prod_{i=1}^n \OO_i (z_i)\ra_{R(t=1)}= \la \prod_{i=m+1}^n \OO_i
(z_i)\ra_{R_2} f(z_{1}, \ldots z_m)\, .
\ee
Combining eqs.\refb{econs2} and \refb{econs3}, and normalizing the
correlator so that the $\la {\bf 1}\ra_\Sigma=1$ on any disk $\Sigma$, we
get
\be \label{econs4}
\la \prod_{i=1}^n \OO_i (z_i)\ra_{R(t=1)}=\la \prod_{i=1}^m \OO_i
(z_i)\ra_{R_1}\,\, \la
\prod_{i=m+1}^n \OO_i
(z_i)\ra_{R_2} \, .
\ee
Thus the correlation function on the splitting surface factors 
into the product
of   correlation functions of the separate surfaces.

\sectiono{Split wave-functionals and Half-string States}   
\label{s3a}

In this section 
we will show that  the  
split wave-functional property of surface states holds when
the boundary of the surface reaches the open
string midpoint.  This
is a very general statement, and our purpose here
will be to explain it just based on the conformal
mapping properties of pinching surfaces, and the factorization
properties of CFT correlators,
both of which were discussed in the previous sections. In the process
we shall find an explicit surface state construction of half-string states
that emerge from the split wave-functional.  
Throughout this section we shall focus on the matter part of the surface 
state only.  

\subsection{Factorization of the string wave-functional} \label{s3a.1}

\newcommand{\DD} {{ \cal D}}

We shall examine the condition under which a string state
$|\Psi\rangle$ 
gives rise to a wave-functional
$\Psi[X(\sigma)]$ that factorizes into a
functional of the left half of the string 
and a functional of the right half of
the string. 
As is clear from the discussion of section \ref{s3.1}, 
all information about the
wave-functional associated to a state 
$|\Psi\ra$  is contained in correlation
functions of the form
\be \label{y1}
\la \Psi | \prod_{i=1}^n \OO_i(\xi_i) |\Psi\ra
= \int [\DD X(\sigma)] \prod_{i=1}^n \widetilde\OO_i(X(\sigma_i)) \Psi[X(\pi
-\sigma)]
\Psi[X(\sigma)]
\, ,
\ee
where $\xi_i = e^{i\sigma_i}$,  $\OO_i$ denote 
an arbitrary set of local vertex
operators, and $\widetilde\OO_i$ 
are these vertex operators viewed as classical
functionals of $X(\sigma)$. 
Let us consider the case where $\sigma_i$ for $1\le i\le m$ lie in the
range $[0, \pi/2)$, and  $\sigma_i$ for $(m+1)\le i\le n$ lie in the range
$(\pi/2, \pi]$.
If the wave-functional is factorized into a functional $\Phi_L$ of the
coordinates of the left-half of the string ($X(\sigma)$ for $0\le\sigma<
\pi/2$) and a functional $\Phi_R$ of the coordinates of the right-half of 
the
string ($X(\sigma)$ for $\pi/2 <\sigma\le \pi$):
\be \label{efactorized}  
\Psi[X(\sigma)]= \Phi_L[X(2\sigma)] \Phi_R[X(2(\pi-\sigma))]\, .
\ee
Note that the parametrization 
of the right half-string has reversed
direction; as $\sigma$ increases we move towards the full string midpoint,
just as for the left half-string.
It now follows that the correlation
function \refb{y1} has the factorized form:
\be \label{y2}
\la \Psi | \prod_{i=1}^n \OO_i(\xi_i) |\Psi\ra
=
f(\sigma_1, \ldots \sigma_m) \, g(\sigma_{m+1}, \ldots \sigma_n)\, .
\ee
Alternatively, if  eq.\refb{y2} is satisfied by all such correlation
functions, we can conclude that $\Psi$ has a factorized wave-functional.
This will be our test of factorization.

Furthermore, if $|\Phi_L\ra$ and $|\Phi_R\ra$ denote the states associated 
with the left and the right half of the string respectively, 
then we have:  
\be \label{y3}
f(\sigma_1, \ldots \sigma_m) = \la \Phi_R^c| \prod_{i=1}^m s\circ
\OO_i(\xi_i) \, |\Phi_L\ra \, ,
\ee
where $s$ denotes the conformal transformation $s: \xi \to \xi^2$ and the 
superscript $c$ denotes twist transformation: $\sigma\to(\pi-\sigma)$, needed
for the right half-string. 
Here the conformal transformation $s$ rescales the
coordinate $\sigma$ so that the coordinate labeling the left half-string
runs from 0 to $\pi$. Similarly 
we have 
\be \label{y4}
g(\sigma_{m+1}, \ldots \sigma_n) = \la \Phi_R^c| \prod_{i=m+1}^n \tilde 
s\circ
\OO_i(\xi_i) \,|\Phi_L\ra
\, ,
\ee
where $\tilde s$ denotes the conformal transformation
$\xi\to \xi^{-2}$. 

\begin{figure}[!ht]
\leavevmode
\begin{center}
\epsfysize=5cm
\epsfbox{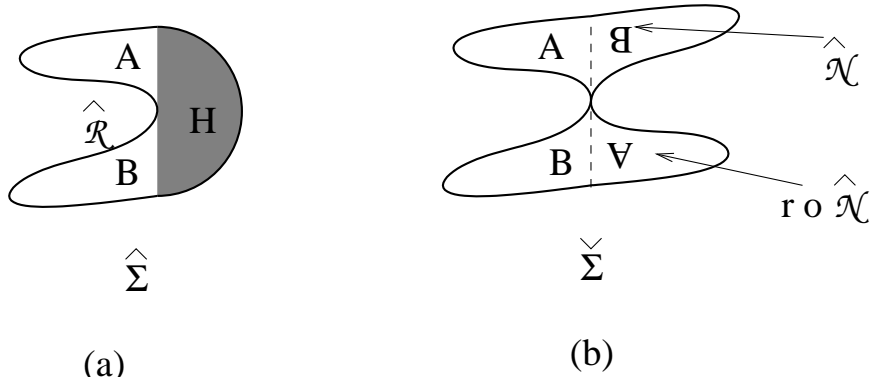}
\end{center}
\caption{The geometry of the disks $\wh\Sigma$ and
$\wc\Sigma$ when the
boundary of $\wh\Sigma$ touches the string midpoint.} \label{fx2}
\end{figure}

We shall now show that \refb{y2} is satisfied for $|\Psi\ra=|\Sigma\ra$ if 
any
part of the boundary of
$\RR$, -- the part of the disk $\Sigma$ outside the local coordinate patch -- 
touches the point $\xi=i$, or equivalently, if in the $\wh w$ plane
any part of the boundary of $\wh \RR$ 
touches the point $\wh w=0$. Such a
situation has been shown in Fig.\ref{fx2}(a). According to the general 
result 
discussed in section \ref{sinner}, the left hand side of 
\refb{y2} for $|\Psi\ra=|\Sigma\ra$ is expressed as a correlation function 
on a surface $\wh\SSS$, obtained by gluing together two copies of 
$\wh\RR$ along the procedure illustrated in Fig.\ref{fx1}. 
In the present context, the gluing of two such
disks
produces a disk
$\wc\Sigma$
which is pinched at the origin of the $\wh w$ plane, as
shown in
Fig.\ref{fx2}(b). In the diagram the images of the points $\sigma_1, 
\ldots
\sigma_m$, lying on the left half-string, are on the positive imaginary
axis, whereas those of the points $\sigma_{m+1},\ldots \sigma_n$, lying on
the right half-string, are on the negative imaginary axis.
$\wc\Sigma$
can be viewed as the union of two disks $\wh\NN$ and $r\circ
\wh \NN$,
joined at the origin, where $r$ denotes the conformal map $\wh w\to -\wh
w$. Since the total surface is pinched, 
the conformal field theory results of section \ref{s2.3} hold, 
and working with normalized correlation functions
so that the partition function
on a disk equals one, 
the correlation function \refb{xx2}
factorizes as
\be \label{xx3}
\la\prod_{i=1}^n h\circ \OO(\xi_i)\ra_{\wc \Sigma}\,
=
\la\prod_{i=1}^m h\circ \OO(\xi_i)\ra_{\wh \NN} \, \,
\la\prod_{i=m+1}^n h\circ \OO(\xi_i)\ra_{r\circ \wh \NN}\, .
\ee
This establishes eq.\refb{y2}. Furthermore this gives:
\be \label{xx4}
f(\sigma_1, \ldots \sigma_m) = \la\prod_{i=1}^m h\circ \OO(\xi_i)\ra_{\wh
\NN}\, .
\ee
Comparing with eq.\refb{y3} we have:
\be \label{xx6}
\la \Phi_R^c| \prod_{i=1}^m s\circ
\OO_i(\xi_i) \, |\Phi_L\ra =    
\la\prod_{i=1}^m h\circ \OO_i(\xi_i)\ra_{\wh
\NN}\,
= \la\prod_{i=1}^m s\circ
\OO_i(\xi_i)\ra_{s\circ h^{-1}\circ\wh
\NN}\, .
\ee
where in the last step we used the conformal
invariance of the correlator to act on the region $\wh\NN$
first by the $h^{-1}$ conformal map, and then by $s$. If the
region $s\circ h^{-1}\circ\wh
\NN$ is simple enough  the explicit identification
of the half string state is possible.

\begin{figure}[!ht]
\leavevmode
\begin{center}
\epsfysize=6cm  
\epsfbox{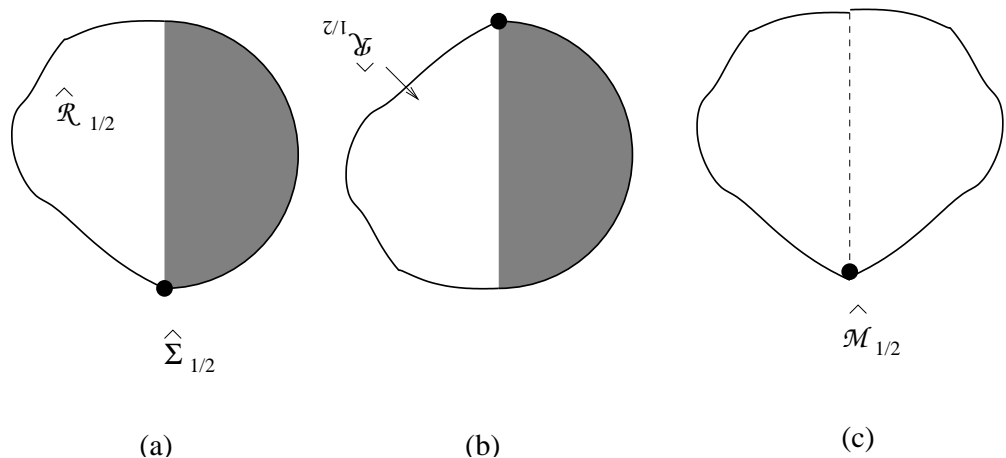}
\end{center}
\caption{Fig. (a) shows the the disk associated with a 
a surface state, describing the state of the half-string, in the $\wh w$ 
plane. Fig.(b) shows the twist conjugate of the surface state of Fig.(a). 
Fig.(c) shows the result of computing inner product between these two 
stares. $\wh\MM_{1/2}$ in this figure is the union of $\wh 
R_{1/2}$ with its  image  under a reflection about the imaginary axis.
The dots signal the half-string endpoint corresponding to the  
full-string midpoint. The other half- string endpoint 
coincides with one of the end-points of the original string.}
\label{ftwist}
\end{figure}

\subsection{Half-string surface states}  \label{s3a.1p}

The above results 
lead to a representation of the state of the half-string as a 
surface state. 
For convenience, let us 
restrict ourselves to the 
case where the original projector was twist invariant, which in this 
context means that $\wh\Sigma$ (and hence $\wh
\RR$) are symmetric under reflection about the real axis. Thus for these
states $|\Phi_L\ra = |\Phi_R\ra\equiv |\Phi\ra$. In that case we can 
rewrite eq.\refb{xx6} as
\be \label{xx6a}
\la \Phi^c| \prod_{i=1}^m s\circ
\OO_i(\xi_i) \, |\Phi\ra  
=\la\prod_{i=1}^m s\circ
\OO_i(\xi_i)\ra_{s\circ h^{-1}\circ\wh
\NN}\, .
\ee
Let us 
take as a trial 
solution for $\la\Phi^c|$ a surface state, represented by a disk 
$\wh\Sigma_{1/2}$ in the $\wh w$ coordinate system, so that
\be \label{ftwist0}
\la\Phi^c|\phi\ra = \la h\circ\phi(0)\ra_{\wh\Sigma_{1/2}}\, .
\ee
As usual we 
denote by 
$\wh \RR_{1/2}$ the part of $\wh\Sigma_{1/2}$ with local coordinate patch 
removed. This has been shown in Fig.\ref{ftwist}(a). The surface state 
associated with $\la\Phi|$ will have an associated $\wh \RR$ which is the 
reflection of $\wh \RR_{1/2}$ about the real axis. This has been shown in 
Fig.\ref{ftwist}(b). In this case, we can 
represent the left hand side of \refb{xx6a} as
\be \label{etwistf1}
\la \Phi^c| \prod_{i=1}^m s\circ
\OO_i(\xi_i) \, |\Phi\ra 
=\la\prod_{i=1}^m h\circ s\circ
\OO_i(\xi_i)\ra_{\wh\MM_{1/2}}\, ,
\ee
where $\wh\MM_{1/2}$ represents the disk obtained by the union of $\wh 
\RR_{1/2}$ with its image under a reflection about the imaginary axis. 
This 
has been shown in Fig.\ref{ftwist}(c). The operators $h\circ s\circ
\OO_i(\xi_i)$ are inserted on the dotted line in this figure. We can 
rewrite this equation as
\be \label{etwistf2}
\la \Phi^c| \prod_{i=1}^m s\circ
\OO_i(\xi_i) \, |\Phi\ra
=\la\prod_{i=1}^m s\circ
\OO_i(\xi_i)\ra_{h^{-1}\circ \wh\MM_{1/2}}\, .
\ee
Comparing \refb{xx6a} and \refb{etwistf2} we get
\be \label{etwistf3}
\wh\MM_{1/2} = h\circ s\circ h^{-1}\circ\wh
\NN\, .
\ee

Let $\wh\RR_{top}$ denote the top wing of $\wh \RR$ associated with the 
original projector describing a state of the full string. This is 
what has been labeled as the 
region $A$ in Fig.\ref{fx2}(a). Given that for twist invariant state the 
region $B$ in Fig.\ref{fx2}(a) is related to the region $A$ by a 
reflection 
about the real 
axis, we see from Fig.\ref{fx2}(b) that the region 
$\wh\NN$ is the union of $\wh \RR_{top}$ with its reflection about the 
imaginary axis. On the other hand we have already seen that $\wh\MM_{1/2}$ 
is the result of the union of $\wh \RR_{1/2}$ with its reflection $I$ 
about the  imaginary axis.
 Finally, it can be easily seen that conjugation by $h\circ 
s\circ h^{-1}$ leaves invariant the  reflection $I$. Thus \refb{etwistf3} 
implies that:
\be\label{toptohalf}
\wh\RR_{1/2} = h \circ s\circ h^{-1} \,\circ \wh\RR_{top} \,.
\ee
Conversely
\be\label{halftotop}
\wh\RR_{top} = h \circ s^{-1} \circ h^{-1} \,\circ\wh\RR_{1/2}\, .
\ee
These equations show how to pass back and forth 
from the split full-string 
surface state to the associated half-string surface state.
We will use this strategy  
to identify the half-string state associated to the butterfly.

\medskip
Before concluding this section we would like to  
explain an issue concerning
boundary conditions. From Fig.\ref{fx2}(a) we see that other than 
the part of the boundary representing the string, the boundary 
of the region $A$ (called $\wh \RR_{top}$ in the current discussion) has 
boundary condition identical to that of the original disk $\wh\Sigma$, 
since 
this part of the boundary of $A$ comes from part of the boundary of 
$\wh\Sigma$. 
Eq.\refb{toptohalf} then implies that the disk $\wh \RR_{1/2}$ also has 
the 
same boundary condition. Thus the half-string state is given by 
the surface state associated with the disk $\wh\Sigma_{1/2}$ in the $\wh 
w$-coordinate system, with 
the same boundary condition as that in the original full string
surface state.

\medskip
This, however, does not imply that both the half-string 
end-points satisfy the 
same boundary conditions as the end-points of the original string, since,
as we 
now explain, 
the surface states {\it defined} 
with certain
boundary conditions may actually fail to {\it satisfy} 
the boundary
condition. 
In a surface state ``the open string" is a specific line with
endpoints at the boundary -- 
in the $\wh w$ presentation 
it is
the vertical boundary of the shaded region called~H. Consider
a boundary condition of Neumann type. This is the statement that
the normal derivative of fields at the boundary vanishes.
We may thus expect on a surface state to find that the expectation
values of the operator 
$\partial_\sigma { X}(\sigma)$  
vanish at $\sigma=0,\pi$.  
But this vanishing will only happen if the
tangent to the open string at the boundary coincides with the
normal derivative to the boundary. This need not be the
case when the boundary has  corners at the open string endpoints.
Corners at the open string endpoints 
happen when the map $z= f(\xi)$ has singularities at the points
$\xi = \pm 1$. While this is not the case for slivers
nor butterflies, we will see examples of this phenomenon 
in section \ref{opr}.

\medskip  
In the case of half-string states associated to projectors,
the above subtleties are in fact quite generic.  If the original
projector is such that the tangent to the open string coincides
with the normal to the boundary, the corresponding half-string
endpoint will carry the boundary condition (this is 
the case illustrated in figure \ref{ftwist}). On the other hand
the nature of the boundary near the full string midpoint, which is controlled
by the behavior of $f(\xi)$ near $\xi = i$, will tell whether or
not the boundary condition is satisfied at the other half-string
endpoint. Getting a little ahead of ourselves we can have a look
at the butterfly state, in particular at figure \ref{ff2}(d).
Note that the tangent to the half-string AQ at Q is indeed along
the normal to the boundary DQ at Q.  Thus we may expect the half-string
state for the butterfly to satisfy Neumann boundary conditions at
both endpoints. It does, because as it will be checked, the half string
state is simply the vacuum state.  On the other hand, for a generalized
butterfly, such as that shown in figure \ref{f2} the two directions
do not coincide and we do not expect the half-string state to satisfy
a simple boundary condition.  For the nothing state, shown in 
figure \ref{fnothing}, the normal to the boundary at the midpoint
is orthogonal to the open string tangent.  Thus the interpretation
here is that $\partial_\tau \wh X$ vanishes at this point.  This is
a Dirichlet boundary condition.

The above considerations may be relevant 
for understanding 
the applicability of the  two different half-string 
formalisms \cite{0202030}, $-$ one where the mid-point satisfies Dirichlet 
boundary condition \cite{halfd} and the other where the midpoint satisfies 
Neumann boundary condition \cite{halfn}. We now see that for half-string
states arising from projectors, neither formalism is natural in all
cases. Since the set of functions are (at least formally) complete,  
we can use either formalism, but the results will take simplest
forms when the boundary conditions match those that arise geometrically
at the string midpoint.

\sectiono{Star Algebra Projectors}\label{new5}  

In this section 
we will show that  the projection  property of surface states
also holds when
the boundary of the surface reaches the open
string midpoint.  We will explain why the projectors 
that arise are of rank one.
Finally we will prove that the Neumann matrix of any
projector has a common eigenvector, the $-1/3$ eigenvector
of the star algebra Neumann matrices.  An intuitive explanation
for this fact is given.

\subsection{Projection properties} \label{s3a.2}

\begin{figure}[!ht]
\leavevmode
\begin{center}
\epsfysize=5cm
\epsfbox{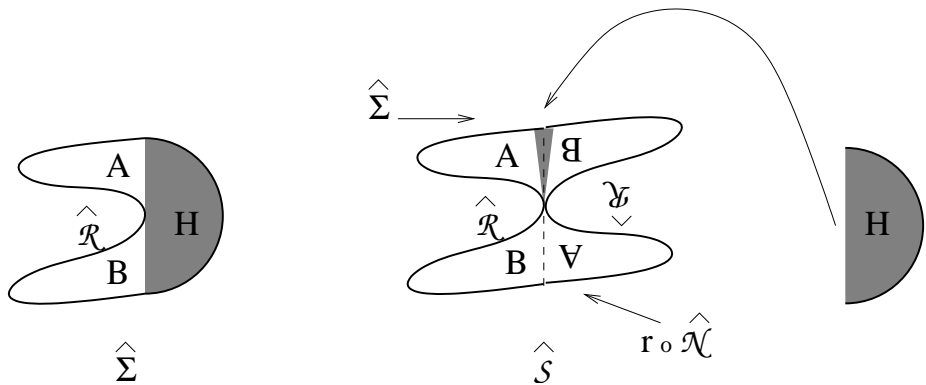}
\end{center}
\caption{The geometry of $\wh\SSS$ when the boundary of $\wh\Sigma$
touches the string midpoint. The local coordinate patch, shown to the
right by the shaded half disk, is to be glued to the shaded region of the
diagram representing $\wh\SSS$.}
\label{fx4} \end{figure}

A surface state $|\Sigma\rangle$ defined as in \refb{yx1} will be called a
projector if it
satisfies:
\be \label{x1}
|\Sigma * \Sigma\rangle = |\Sigma\rangle\, .
\ee
We shall show that a surface states $|\Sigma\rangle$ is a projector if the
corresponding surface $\Sigma$ has the property that the boundary of
$\Sigma$ touches the string mid-point $\xi=i$, -- the same condition under
which its wave-functional factorizes.
However, unlike in the previous section, in this subsection 
we shall work with the full surface state in the matter-ghost conformal 
field theory so that the total central charge vanishes, and the gluing 
relations required for the computation of $*$-product
are valid without any additional multiplicative factors.
For computing $|\Sigma*\Sigma\ra$ we follow the procedure described in 
section \ref{sinner}.
In this case the surface $\wh\SSS$ that appears in eq.\refb{estar}, 
constructed following Fig.\ref{fx3}, 
is the pinched union of
$\wh\Sigma$ and an extra disk $r\circ\wh\NN$  
as shown in Fig.\ref{fx4}. We have, 
as in eq.\refb{estar},
\be \label{estara}
\la \Sigma*\Sigma|\phi\ra = \la h\circ\phi(0)\ra_{\wh\SSS}\, .
\ee
The operator
$h\circ\phi(0)$ is being inserted on the  
$\wh\Sigma$ component of $\wh\SSS$,  
on the boundary of $H$, as usual.  
Hence there is no operator insertion on $r\circ\wh\NN$.  
Our factorization result of section \ref{s2.3} implies that the correlator
factorizes and the contribution of $r\circ\wh\NN$   
is simply a multiplicative
factor of one. Thus eq.\refb{estara} can be rewritten as:
\be \label{x9}
\la \Sigma*\Sigma|\phi\ra = \la h\circ \phi(0)\ra_{\wh\Sigma}\, .
\ee
But the above right hand side is precisely $\la\Sigma|\phi\ra$
and therefore  this 
establishes \refb{x1}.

\medskip 
There is also a nice geometrical understanding that
projectors that arise as surface states of the type discussed
above are of rank one, -- at least in a limited sense.  
For  operators on separable
Hilbert spaces, a projector $P$ is of rank one if and 
only if $ P A P = \hbox{Tr} (AP)  \, P$ for all $A$.  Let now $\Sigma$ be
a surface state projector $\Sigma * \Sigma = \Sigma$, and let
$\Upsilon$ denote an arbitrary state of the star algebra
(a Fock space state, or a surface state, for example).
We then claim that the condition characterizing $\Sigma$ as a rank
one projector holds:
\be
\label{rone}
\Sigma * \Upsilon * \Sigma =  \langle \Sigma | \Upsilon\rangle \, \Sigma \,.
\ee
This equation is understandable in terms of pictures. Back to 
Fig.\ref{fx4}, the above left hand side would be represented by
a modified $\wh\SSS$ where the $r\circ \wh\NN$ 
disk would be changed
by cutting open the dashed line separating the sides $B$ and $A$, and
gluing in the state $\Upsilon$. The new $r\circ \wh\NN$ 
disk, still pinched
with respect to the remaining surface $\wh\Sigma$ would be
producing the $ \langle \Sigma | \Upsilon\rangle $ inner product. 
The factorization implied by the pinching of the surfaces then
yields \refb{rone}.

There is, however, a subtlety involved in the derivation of \refb{rone} 
which we now discuss. In order to apply the factorization results of 
section \ref{s2.3} we need a unitary BCFT -- otherwise the contribution from 
operators of 
negative dimension to the operator product expansion will invalidate 
\refb{econs4}. Thus \refb{rone} is not valid in general for the combined 
matter-ghost system, but could be valid for example for the matter 
part of 
the surface state. On the other hand since the matter part of the BCFT has 
a non-zero central charge, gluing of surface states typically involve 
(possibly infinite) multiplicative factors. Thus we expect \refb{rone} to 
be valid for the matter part of the state up to an overall multiplicative 
factor that depends only on the central charge of the matter BCFT and does 
not depend on the state $|\Upsilon\ra$ or the particular BCFT under 
consideration. Alternatively, \refb{rone} is  valid without any additional 
multiplicative factor in the combined matter-ghost BCFT {\it if we 
restrict $|\Upsilon\ra$ to be a state of ghost number 0}, so that the 
leading contribution to the factorization relation comes from the identity 
operator as has been assumed in the derivation of \refb{econs4}. 

\subsection{A universal eigenvector of $V^f$
for all projectors}   

Give a projector of the type described above, we can use the procedure of 
section \ref{soper} to represent it as
the exponentials of matter 
(and ghost) 
oscillators acting on the vacuum as in \refb{soscill}. We shall now show 
that the matrix $V^f_{mn}$ associated with any projector has the property 
that it has an eigenvector of eigenvalue
one, the eigenvector being 
the same as the $\kappa=0$ 
eigenvector of
$K_1$\cite{0111281}. This generalizes the same property obeyed by the 
sliver 
Neumann matrix.

We start with \refb{evmn} and integrate
by parts with respect to $w$,
\ben
\label{vsimp}
V^f_{mn} = {(-1)^{m} \over \sqrt{m}}
\oint {dw\over 2\pi i} \oint {dz\over 2\pi i} {1\over z^m}
{f'(z)\over
f(z) - f(w)} \,\sqrt{n} \, \Bigl( -{1\over w}\Bigr)^{n+1} \,.
\een
For definiteness we shall take the $w$ contour to be outside the $z$
contour.
Acting on an eigenvector with components $v_n$ we find
\ben
\label{simple}
\sum_{n=1}^\infty V^f_{mn}\, v_n = {(-1)^{m} \over \sqrt{m}}
\oint {dw\over 2\pi i}
\oint {dz\over 2\pi i} {1\over z^m}
{f'(z)\over
f(z) - f(w)}\,\,{1\over w^2}\sum_{n=1}^\infty
\sqrt{n}\, v_n    \Bigl( -{1\over w}\Bigr)^{n-1} \,.
\een

\bigskip
We have argued that for all sufficently well-behaved
maps $f(\xi)$
such that $f(\pm i) = \infty$, $\Psi_f$ is
a projector. We now show that  $f(\pm i) = \infty$
suffices to show that the
$C$-odd $\kappa=0$ eigenvector $v^-$ of the Neumann
matrices is in fact an eigenvector of eigenvalue
one for the matrix $V^f$. The eigenvector in question is
defined by the generating function
\be
\label{genv}
\sum_{n=1}^\infty  {v_n^-\over \sqrt{n}}\, u^n =
\tan^{-1} u
\quad \to
\quad
\sum_{n=1}^\infty \sqrt{n} v_n^- \, u^{n-1} = {1\over 1+ u^2}\,\qquad
|u| <1.
\ee
For regulation purposes we pick a
number $a$ slightly bigger than one and write
\be
\label{regv}
\sum_{n=1}^\infty \sqrt{n} v_n^- \, u^{n-1} = {1\over  1 +
(u/a)^2}\, , \qquad
|u| < a \, ,
\ee
with the  understanding that
the limit $a\to 1^+$ is
to be taken.
Therefore back in \refb{simple}  we get
\ben
\label{simpleonv}
\sum_{n=1}^\infty V_{mn}\, v_n^- = {(-1)^{m} \over \sqrt{m}}
\oint {dw\over 2\pi i}
\oint {dz\over 2\pi i} {1\over z^m}
{f'(z)\over
f(z) - f(w)}\,\,{a^2\over 1+a^2 w^2} \,, \quad  |w| > \frac{1}{a}.
\een
The $w$ integral must run over a contour of radius bigger than
$1/a$ because of the use of \refb{genv} with $w= -1/u$.
At the same time the radius of the contour must be less than
1 so that the contour does not enclose the singularities at $w=\pm i$.
Therefore we pick  up contributions from the poles at $w= \pm i/a$ and
$w=z$.
After this we can take the $a\to 1$ limit.
Since  $f(\pm i) = \infty$,  only the $w=z$
pole contributes and we get
\ben
\sum_n V_{mn} v_n^-
=  {(-1)^{m+1}\over \sqrt{ m}}
\oint {dz\over 2\pi i} {1\over z^m} {1\over 1+ z^2}
= v_m^-\, .
\een
This establishes the claim.

An intuitive explanation for this property can be given using
the half-string interpretation. The eigenvector in question implies
that the projector wave-functional is invariant under constant
and opposite translation of the half-strings \cite{0111069}.
If we denote by $P_L$ and $P_R$ respectively the momentum carried
by the left and right half-strings we have that the eigenvector
condition is interpreted as the condition that
\be
(P_L - P_R) | \Sigma\rangle =0\,,
\ee
where $|\Sigma\rangle$ is the projector surface state.  Being a 
surface state defined with Neumann boundary condition 
(in the sense described in section \ref{s3a.1p}), 
the total momentum $P_L+ P_R$ carried by the state vanishes.
Thus the condition above is simply the statement that $P_L$ {\it and}
$P_R$ annihilate $|\Sigma\ra$. But this must be so, 
since 
$|\Sigma\ra = |\Sigma_L\ra
\otimes |\Sigma_R\ra$, 
where $|\Sigma_L\ra$ and $|\Sigma_R\ra$ are themselves
surface states defined with Neumann boundary condition.

\sectiono{The Butterfly State} \label{s2b}

In this section we shall introduce and investigate in
detail the butterfly surface state.  After giving the details
of its definition and viewing it in various possible ways
we shall verify that it is a projector of the star  algebra.
Throughout this section
except in section \ref{sbutterwave}
we shall work
in  the
combined matter-ghost
system with zero central charge so that we can  apply the gluing
theorem without any additional factors coming from  conformal anomaly.

\subsection{A picture of the butterfly} \label{s2a.1}

\begin{figure}[!ht]
\leavevmode
\begin{center}
\epsfysize=14cm
\epsfbox{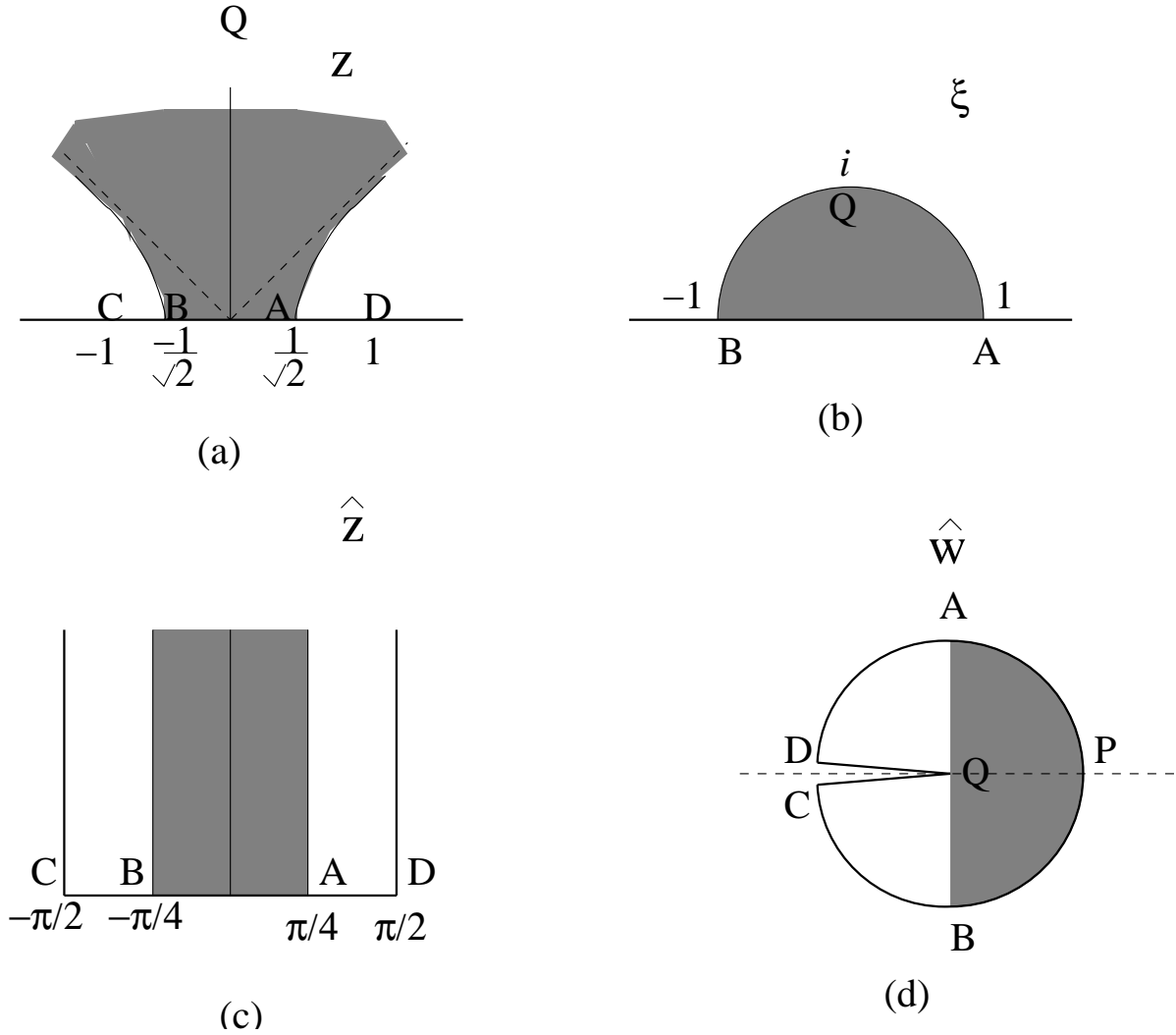}
\end{center}
\caption{Representation of the disk associated with the butterfly state
in various coordinate systems.
The shaded region denotes the
local coordinate patch.} \label{ff2}
\end{figure}

The butterfly state, just as any surface state, is completely
defined by a map from $\xi$ to the upper half $z$ plane
($z$-presentation, as reviewed in section 2.1).  We thus
write
\be
\label{defbut}
z = {\xi\over \sqrt{1+ \xi^2}}\equiv f_{\BB}(\xi)\, ,
\ee
and define the butterfly state $|\BB\ra$ through the relation:
\be \label{ebutterdef}
\la\BB|\phi\ra = \la f_\BB\circ\phi(0)\ra_{UHP}\, .
\ee
In the $z$-presentation the surface is the full upper half
plane, and therefore in order to gain
intuition about the type of state this is, we plot the image
of the canonical $\xi$ half-disk in the $z$-plane (see Fig. \ref{ff2} (a)
and
(b)). The open string $|\xi| =1, \Im (\xi) \ge 0$ is seen to map to the
hyperbola
$x^2-y^2 = {1\over 2}$ (in the upper half plane, with $z= x+ iy$).
We note that $z(\xi=i) = \infty$ and thus, as expected for a
projector,  
the open string midpoint coincides with the boundary
of the disk.

Further insight into the nature of the state is obtained
by examination of the disk in the $\wh z$-presentation.
To this end we use \refb{xx22} to recognize that \refb{defbut}
can be rewritten as
\be
\label{moddef}
z = \sin (\tan^{-1}(\xi)) = \sin \wh z
\ee
This maps the image of the local coordinate in
the $\wh z$-presentation to the image of the local
coordinate in the $z$-presentation.  As explained before,
the surface need not fill the upper-half $\wh z$-plane.
To figure out the extension of the surface in the $\wh z$ presentation
we simply invert the previous equation to write
\be
\label{moddeff}
\wh z = \sin^{-1} z \,.
\ee
As shown in the figure (\ref{ff2} (c)), this transformation maps the full
upper
half $z$-plane into the region $|\Re (\wh z)| \leq \pi/2\,, \Im (\wh z)
\geq 0$. Note that the vertical lines $\Re (\wh z) = \pm\pi/2$
are images of the boundary and not identification lines.  Even though
the surface occupies a portion of the $\wh z$-plane the boundary
reaches the point at infinity, and so does the midpoint (as expected).
The above conformal map is perhaps most easily thought about
in differential form, where it belongs to the class of
Schwarz-Christoffel transformations.  We have
\be
\label{difform}
d\wh z = {dz\over \sqrt{(1-z)(1+z)}}
\ee
The real line in the $z$-plane is mapped into a polygon in
the $\wh z$ presentation, where the turning points are
$z=\pm 1$ and the turning angles are both $\pi/2$. This, of course
is the result shown in the figure.

Finally, we give the $\wh w$ presentation (fig.~\ref{ff2}(d)). Using
\refb{hwhz}
the region $|\Re (\wh z)| \leq \pi/2\,, \Im (\wh z)
\geq 0$ of the $\wh z$ presentation turns into the full
disk with a pair of cuts zooming into the $\wh w$ origin
from $\wh w = -1$.
Indeed the boundary
of the  surface
is the arc $e^{i\theta}$ with 
$ 0 <\theta < \pi$ 
together with
the line going from $\wh w = -1$ to $\wh w =0$, plus the backwards
line from $\wh w=0$ to $\wh w = -1$ plus the arc
$e^{i\theta}$ with $-\pi <\theta < 0$. 
It is
perhaps in this presentation that it is clearest that the string
midpoint
$\wh w=0$ touches the boundary of the disk.

\subsection{The regulated butterfly} \label{ereg}

\begin{figure}[!ht]
\leavevmode
\begin{center}
\epsfysize=6cm
\epsfbox{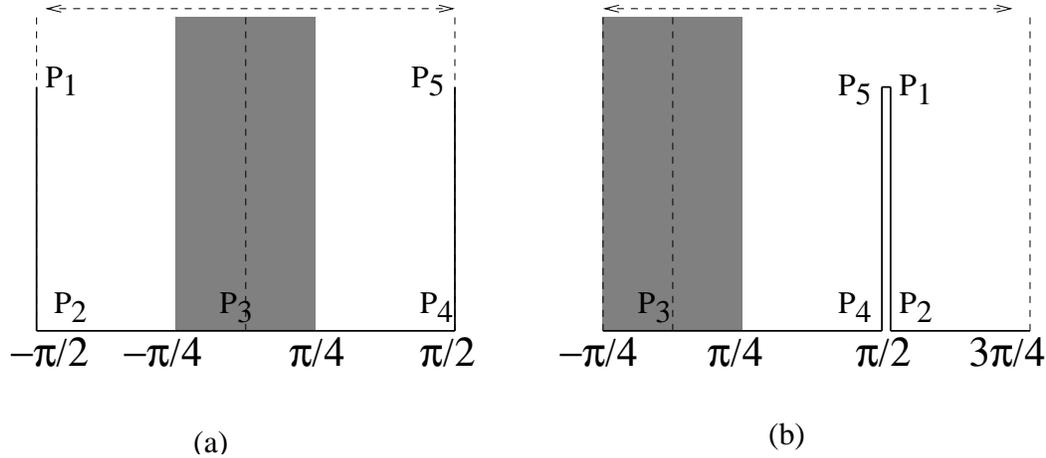}
\end{center}
\caption{The geometry of the disk associated with the
regularized butterfly in the complex $\wh z$ plane.
The shaded region denotes the
local coordinate patch. The lines $\Re(\wh z)=-\pi/4$, $\Re(\wh
z)=3\pi/4$ are identified in the second figure.} \label{ff3}
\end{figure}

In order to regulate the butterfly it is simplest to do it
in the $\wh w$ coordinates. Here we simply stop the cut
at some point $\wh w_0$ with $-1< \wh w_0<0$ a real negative
number. In the $\wh z$- presentation this turns into a
picture shown in fig \ref{ff3} (a). Note that the vertical lines $\Re
(\wh z) =
\pm\pi/2$ are not all boundary. Indeed the segments $\overline{P_1P_2}$
and $\overline{P_4P_5}$ are part of the boundary, but the remaining
parts of the vertical lines, shown dashed in the figure, are
identification lines. For more clarity and also
later convenience we have
shown in fig.
\ref{ff3}(b)
the same regulated butterfly where the region to the left of the
right-half string has been moved, and the two extreme vertical lines
are identified.

\begin{figure}[!ht]
\leavevmode
\begin{center}
\epsfysize=6cm
\epsfbox{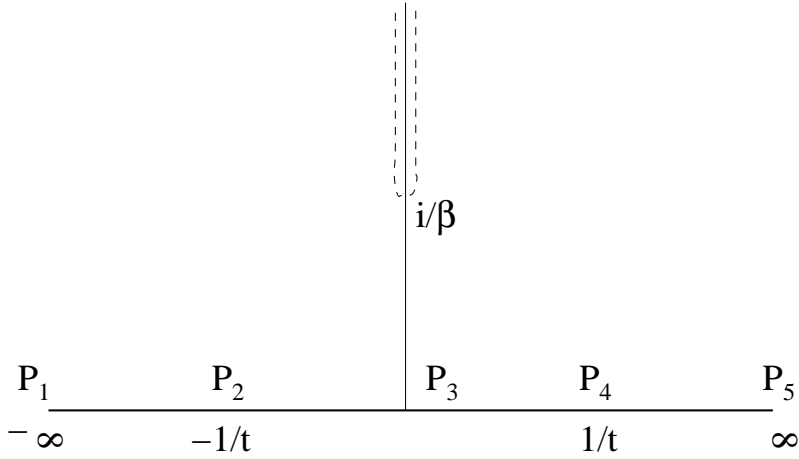}
\end{center}
\caption{The singular points of the map $z\to\wh z$ for the regulated
butterfly in the $z$-plane.}
\label{ff4}
\end{figure}

In order to find the relation between $z$ and $\wh z$ in the
regulated butterfly we must construct the map, which is a variation
on \refb{difform}. 
In order to produce the
identification shown by dashed lines, while preserving the property
that the real line is mapped to a polygon a
pair of complex conjugate poles are necessary.
We write
\be
\label{begmap}
d\wh z = {dz\over (1+ \beta^2 z^2) \sqrt{1- z^2 t^2}}
\ee
where we have fixed the normalization from the condition
${d\wh z\over dz}= 1$ at $z=0$. The images of the marked points
$P_1,\ldots P_5$ in Fig.\ref{ff3}(a) in the $z$-plane is
indicated in figure \ref{ff4}. The identification
lines emerge from the pole at $z= i/\beta$. Since the identification
lines differ by $\Delta \wh z = \pi$ the residue at the pole
$z= i/\beta$ in \refb{begmap}  must equal ${1\over 2i}$. This requires
$\beta^2 = 1-t^2$ and thus we have
\be
\label{begmapp}
d\wh z = {dz\over (1+ (1-t^2) z^2) \sqrt{1- z^2 t^2}} \,.
\ee
This equation is readily integrated to give
\be
\wh z = \tan^{-1} \Bigl( {z\over \sqrt{1-z^2 t^2}} \Bigr) \,,
\ee
and inverting the relation one finds
\be
\label{defrbut}
z =  {\tan \wh z\over \sqrt{1+ t^2 \tan^2 \wh z}} =
 {\xi\over \sqrt{1+ t^2 \xi^2}}\,.
\ee
This is
a rather simple result. The regulator
parameter $t$ can be related to the height $h$ of the
points $P_1$ or $P_5$. Say, for $P_5$, $\wh z (P_5) = {\pi\over 2} +
ih$ must map to $z=\infty$. This requires $1+ t^2 \tan^2 ({\pi\over 2} +
ih) =0$. A short calculation gives
\be
h = \tanh^{-1} t\,, \quad  \quad\hbox{thus} \quad h\to\infty
\quad \hbox{when} \quad t \to 1^- \,.
\ee
The regulator parameter $t$ must therefore satisfy $t<1$.
Clearly when $t=1$ in \refb{defrbut} we recover the butterfly as
defined in \refb{defbut}.

\subsection{Star multiplying two regulated butterflies} \label{sregstar}

\begin{figure}[!ht]
\leavevmode
\begin{center}
\epsfysize=10cm
\epsfbox{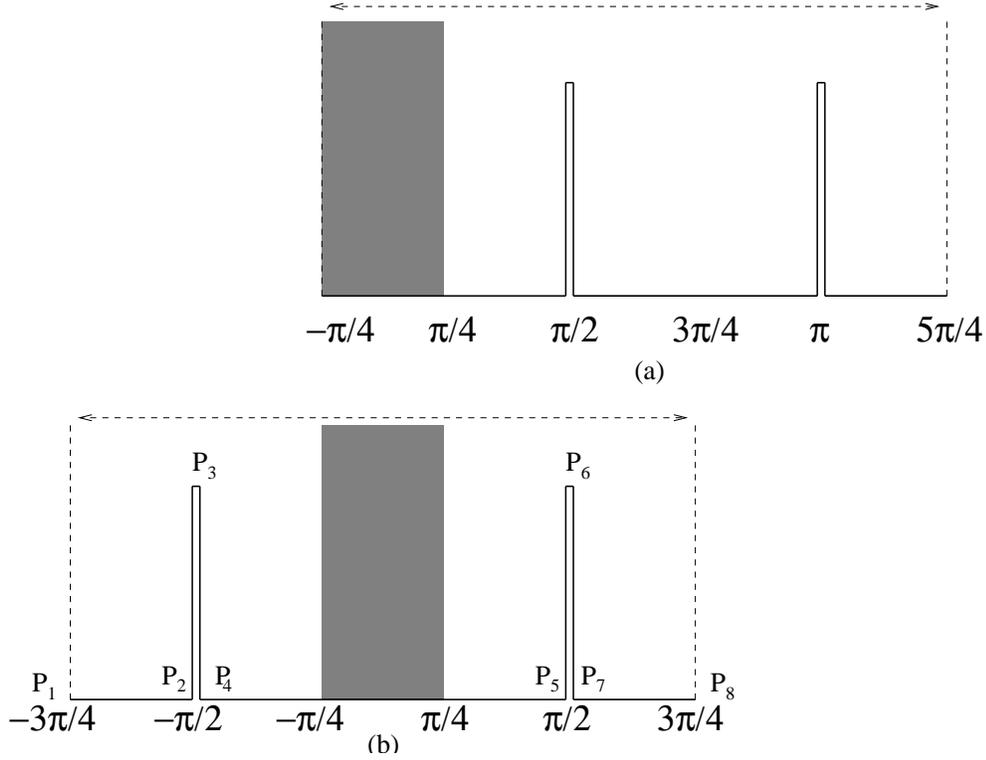}
\end{center}
\caption{Representation of the $*$-product of a regulated butterfly with
itself in the $\wh z$ plane.} \label{ff5}
\end{figure}

\begin{figure}[!ht]
\leavevmode
\begin{center}
\epsfysize=6cm
\epsfbox{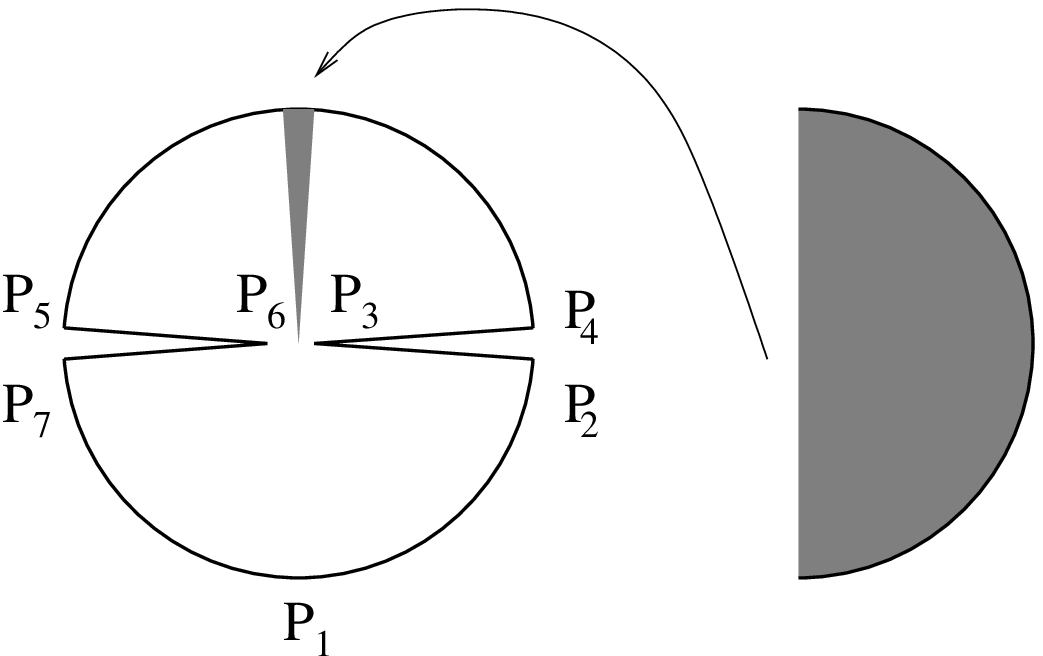}
\end{center}
\caption{Representation of the $*$-product of a regulated butterfly with
itself in the $\wh w$ plane. The local coordinate patch, which is to be
glued to the rest of the digram, is shown separately on the right.}
\label{ff5a} \end{figure}
To star multiply two regulated butterflies we take the first
one, and glue to the right-half of its open string the left-half
of the open string of the second butterfly, whose local coordinate patch has
been removed. In order to perform these operations it is
easier to view the butterfly as the cylinder $-{\pi\over 4}\leq \Re
(\wh z) \leq {3\pi\over 4}, \, \Im (\wh z) \ge 0$ with the
vertical lines, corresponding to the right half of the open string,
identified (see fig. \ref{ff3}(b)).  The second (amputated) butterfly can
be
glued to the right of this one giving the result in fig \ref{ff5}(a).
Finally
we choose a symmetric arrangement of this figure as shown in figure
\ref{ff5}(b).
Special points
$P_1, \cdots P_8$ have been marked,
and the complete picture is a cylinder with circumference ${3\pi/2}$
and with the dashed lines identified. The image of this disk in the $\wh
w$
coordinate system is shown in Fig.\ref{ff5a}. In the $t\to 1$ limit,
the vertices $P_3$ and $P_6$ of the two wedges approach the origin of the
$\wh w$
plane, and in this limit the surface clearly has the stucture  of a split
disk of the form discussed in section \ref{s2.2}.

\begin{figure}[!ht]
\leavevmode
\begin{center}
\epsfysize=5.5cm
\epsfbox{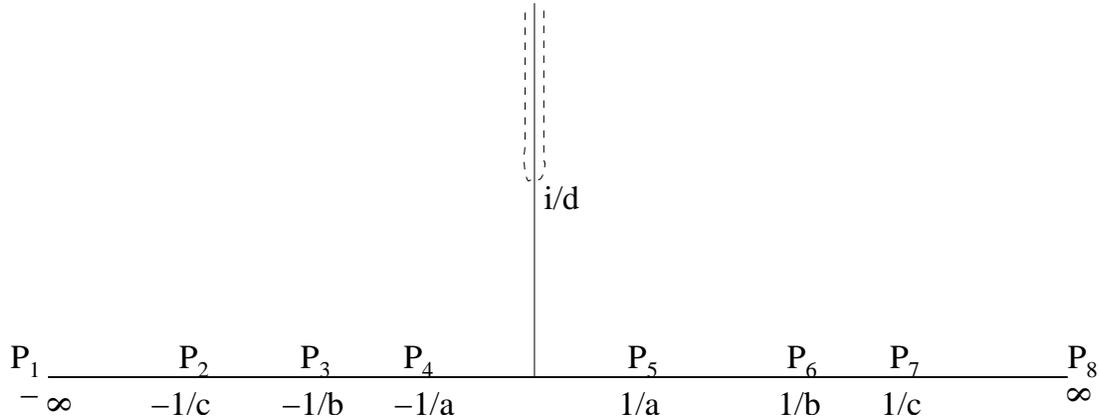}
\end{center}
\caption{The singular points of the map $z\to\wh z$ for the $*$-product of
two regulated butterflies in the $z$-plane.}
\label{ff6}
\end{figure}

The map of this
nontrivial polygon in the $\wh z$ plane into the upper half $z$-plane is
defined by a map whose
singularity structure is symmetrically arranged as shown in figure
\ref{ff6}.
Taking into account the various turning points, the map is of the form
\be
\label{starmap}
d\wh z = {1\over (1+ d^2\, z^2)} \,{ (1 - b^2 z^2) \over \sqrt{
(1- a^2 z^2) (1-c^2 z^2) }}\, dz  \equiv 
H(z) dz\,.
\ee
The complex poles at $z= \pm {i\over d}$ play no role in the
turning points but are needed for the implementation of the
identification of the dashed lines in the $\wh z$-plane (figure
\ref{ff5}(b)).
The length conditions are
\ben
\int_0^{1/a} H(z) dz  &=& {\pi\over 2}\,, \label{con1}\\
\int_{1/a}^{1/b} H(z) dz  &=& ih \,,\label{con2}\\
\int_{1/b}^{1/c} H(z) dz  &=& -ih \,,\label{con3}\\
\int_{1\over c}^{\infty} H(z) dz  &=& {\pi\over 4} \,,\label{con4}
\een
where
\be \label{edefh}
h=\tanh^{-1} t\, .
\ee
These are 
four equations, for our four unknowns $a,b,c$ and $d$. These four
equations, with the analogous ones integrating over the negative $z$ axis,
added together imply that
\be
\label{eqsum}
\int_{-\infty}^{\infty}
H(z) dz = {3\pi\over 2}\,.
\ee
This means that
that the residue around
$z={i\over d}$ in \refb{starmap} must equal ${3\over 4i}$.  A short
calculation shows that this residue condition requires
\be
\label{rescon}
{9\over 4} \, d^2 \,(d^2 + a^2 ) (d^2 + c^2 ) = (d^2 + b^2 )^2 \,.
\ee
It should be noted that this residue condition is not independent
from conditions listed in \refb{con1} to \refb{con4}.

\medskip
The issue to
be examined here is how to achieve very large
$h$ by adjusting the parameters $a,b,c$ and $d$. The analysis that
follows is a special case of the discussion in section 2.2. As the
slits become higher and higher by growing $h$ the surface is pinching. In
the representation of Fig.\ref{ff5}(a), 
the
role of $R_2$ is played by the region ${\pi\over 2}\le \Re(\wh z)\le \pi$, 
$\Im (\wh z) \ge 0$, which in Fig.\ref{ff5}(b) 
corresponds to the region
${\pi\over 2} \leq |\Re(\wh z)|
\leq {3\pi\over 4}\,, \Im (\wh z) \ge 0$. Our expectation is therefore
that as the slits go up to infinity, this region vanishes away in a map
that preserves the inner region, and
we recover a single butterfly.  We will
now show that the large $h$ limit can be achieved by
taking $b,c$ and $d$ much smaller than $a$, and $c$ to be much smaller
than $b,d$:

\be
\label{fon}
\{ b,c,d\} \ll a \,, \qquad c\ll \{b,d\}\, .
\ee
Since such small parameters imply that the turning points
$1/b, 1/c$ and $1/d$ are going to infinity, it is convenient
to bring them near zero to understand how they are collapsing
into each other. We therefore let $z = -1/z'$ and find that
\refb{starmap} gives
\be
\label{starmapp}
d\wh z = {1\over (z'^2+ d^2)} \,{ (z'^2 - b^2 ) \over \sqrt{
(z'^2- a^2 ) (z'^2 -c^2 ) }}\, dz'  \equiv 
G(z') dz'\,.
\ee
Our first condition will be to achieve \refb{con1}. This gives
\be
\int_a^\infty G(z') dz'  \simeq \int_a^\infty {dz'\over z'}
{1\over \sqrt{z'^2 - a^2 } } =  {\pi\over 2} \,,
\ee
where we have noted in \refb{starmapp} that for $a< z'<\infty$ and
the inequalities in \refb{fon}, $z' \gg b,c,d$, and the
expression for $G(z')$
simplifies considerably.  This equation requires
\be
\label{firsta}
a \simeq 1 \,.
\ee
With $a \simeq 1$ and much bigger than $b,c$ and $d$,
and $c$ much smaller
than $b$, $d$,
equation \refb{rescon}
now gives
\be
\label{bandd}
{1\over 2} \, d^2 \simeq b^2 \,.
\ee
With $d$ comparable to $b$, and $c\ll\{d,b\}$
we now claim that all the
conditions
listed in \refb{con1} to \refb{con4}, and the demand that $h$ be large,
can be satisfied.
Since \refb{rescon} and thus \refb{bandd} is a
consequence of \refb{con1} and
\refb{con4}, and we have already satisfied \refb{con1}, we should be able to
see that \refb{con4} is satisfied. Indeed, we must have
\be
\int_0^c G(z') dz' \simeq \int_0^c {b^2\over d^2 \sqrt{c^2 - z'^2}} dz'
\simeq {\pi\over 4}\,.
\ee
A short calculation shows that this equation holds on accound of
\refb{bandd}.

It only remains now to verify that conditions \refb{con2} and \refb{con3}
can be satisfied with ever increasing $h$. Condition\refb{con2} requires
that the integral
\be
\int_b^1 {1\over z'^2 + 2b^2} \,{z'^2 - b^2 \over \sqrt{1-z'^2}}\, {dz'\over
z'}\,
\ee
obtained using \refb{fon}
grow without bounds as $b$ is made
progressively small. This is clearly the case, since the integral
diverges at the bottom limit when $b$ is set to zero.  This means that
for any $h$ we can satisfy \refb{con2}
for sufficiently small $b$,  with $c\ll b$.  Condition \refb{con3}
requires
that the integral
\be
\int_c^b {dz'\over z'^2 + 2b^2} \,{z'^2 - b^2 \over \sqrt{z'^2-c^2}}\,
\,
\ee
obtained using \refb{fon},
grow without bounds as $c$ is made
progressively small. This is clearly the case, since the integral
diverges at the bottom limit when $c$ is set to zero.  This means that
having satisfied \refb{con2} for a fixed and very large $h$ by choosing
a sufficiently small $b$ while keeping $c\ll b$, we can now satisfy
\refb{con3} by making $c$ sufficiently small.

We have thus shown that as we multiply two regulated
butterflies and let the regulator go away, the map defining
the composite surface is that of \refb{starmap}, with the limit
$a\to 1$, and $\{ b,c,d\} \to 0$ taken. This gives us
\be
\label{starmapdone}
d\wh z =  \,{ 1  \over \sqrt{1-  z^2 }}\, dz \,,
\ee
which, by comparison to \refb{difform}, it is immediately
recognized to be the definition of the butterfly.  This concludes
our proof that the butterfly emerges from the star product of
two regulated butterflies in the limit as the regulator is removed.

\medskip
It is natural to wonder if the multiplication of two regulated
butterflies also gives a 
regulated butterfly for $t\simeq 1$. 
We find that this product is a 
butterfly 
regulated in a slightly different way. To see this note that
using \refb{bandd}, we can go back to \refb{rescon} to find a more
accurate evaluation of $a$, which previously was just set to one.
We find
\be
d^2 \simeq 1-a^2\,.
\ee
With such relation, the map \refb{starmap} becomes
\be
\label{starmapreg}
d\wh z = {1\over (1+ (1-a^2)\, z^2)} \,{ (1 - {1\over 2} (1-a^2) z^2) \over
\sqrt{ 1- a^2 z^2 }}\, dz \,,
\ee
where we used $c\ll\{ b,d\}$.  The correspondence with
the regulated butterfly map given in \refb{begmapp} is very
close, but not exact.  
The reason for this is intuitively clear. Our conformal map
statement in section \ref{s2.2} states that the map that shrinks
away the extra surface at the other side of a thin neck only
affects the region around this neck. In addition, 
regulators control the approach of the boundary 
to the open string midpoint. Since
the neck arising from star multiplication occurs around the 
open string midpoint of the product string (see, for example
fig.~\ref{fx4}), the regulator arising
after star product is affected by the way in which the map shrinks
away the extra surface.

\subsection{Half-string wave-functional for  
 the butterfly state} \label{sbutterwave}  
 
\begin{figure}[!ht]
\leavevmode
\begin{center}
\epsfysize=8cm  
\epsfbox{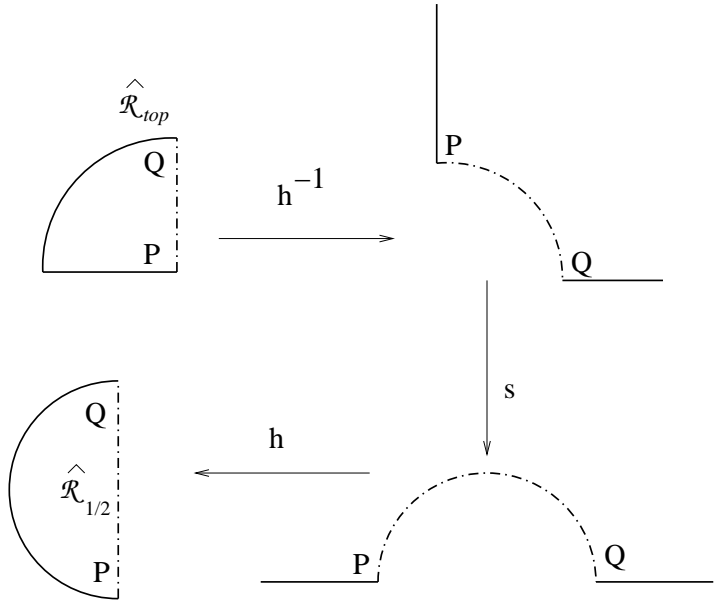}
\end{center}
\caption{The $\wh\RR_{1/2}$ associated with the butterfly state, and 
its images under the maps $h^{-1}$, $s\circ
h^{-1}$ and $h\circ s\circ h^{-1}$. $P$ and $Q$ are two marked points on 
the boundary of the disk, and the labels $P$ and $Q$ are always located 
in the inside of the disk.} \label{f4new}
\end{figure}

\begin{figure}[!ht]
\leavevmode
\begin{center}
\epsfysize=4cm
\epsfbox{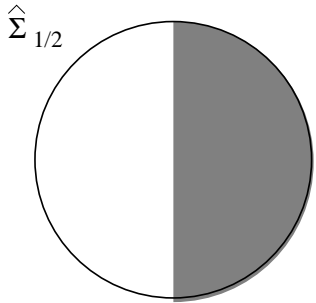}
\end{center}
\caption{The $\wh\Sigma_{1/2}$ associated with the butterfly.}
\label{f5newer}
\end{figure}

According to the general arguments given in section \ref{s3a.1}, the
wave-functional of the butterfly state splits into a product of a
functional of the left half of the string and a functional of the right
half of the string. We can now ask: what particular half-string
wave-functional does the butterfly state have? To answer this question we
go back to eq.\refb{toptohalf}. In the case of the butterfly, 
$\wh\RR_{top}$ is the unit quarter disk in the second quadrant as shown in 
the 
top left hand diagram of
Fig.\ref{f4new}. As shown in the rest of the figure, under the map $h 
\circ 
s\circ h^{-1}$ this gets mapped to the unit half-disk to the left of the 
vertical axis. This then is the $\wh\RR_{1/2}$ for the 
butterfly. Thus the disk $\wh\Sigma_{1/2}$, obtained by joining with 
$\wh\RR_{1/2}$ the copy of the local coordinate patch, is the full unit 
disk as shown in Fig.\ref{f5newer}.
This gives the half string state associated with the butterfly to be:
\be \label{ebutterfn}
|\Phi\ra = |0\ra\, ,
\ee
thereby establishing that the half-string wave-functional associated with
the butterfly state is the vacuum state.

\subsection{Operator representation of the butterfly state} 
\label{sbutterope}

We can represent the regulated butterfly $|\BB_t\ra$ in the operator 
formalism 
following the general procedure outlined in section \ref{soper}. 
In this case we have
\be  z =  f_t(\xi) =  \frac{\xi}{\sqrt{1+ t^2 \xi^2}}=
\exp \Bigl( v_t (\xi) {\partial\over \partial \xi}
\Bigr)\, \xi \,.
\ee
Eqs.\refb{egenc1}, \refb{egenc2} give
\be \label{vbutter}
v_t(\xi) = -t^2  \, \xi^3 /2 \, .
\ee
Eq.\refb{Uf}, \refb{vdef} now gives:
\be \label{Bexp}
|\BB_{\,t} \ra = \exp \left( -\frac{t^2}{2} \,L_{-2} \right)|0\ra\,.
\ee
This is a remarkably simple expression involving a single Virasoro
operator in the exponent.

The formalism of Virasoro conservation laws \cite{0006240}
allows us 
to derive an interesting property of the butterfly state,
\be \label{PK}
K_{2} |\BB \rangle =0 \, ,
\ee
where $K_{2} = L_{2} -  L_{-2}$.
Indeed, consider in the global UHP 
the vector field
\be
\tilde v_{2}(z)=2 z - \frac{1}{z} \, ,
\ee
which is holomorphic everywhere, including infinity,
except for the pole at the puncture $z=0$. It follows
that
\be  
\Big\la \oint dz \;  \tilde T(z) \,\tilde v_{2}(z) 
f_{\BB} \circ\phi(0)\Big \ra_{UHP} = 0 \, ,
\ee
for any state $|\phi\ra$,
where the contour circles the origin.
Changing variables to the local coordinate $\xi$, 
we find
\be  
\Big\la f_{\BB} \circ \Big( \oint d\xi \; T(\xi)   
(\xi^{3}-\xi^{-1}) 
\phi(0) \Big) \Big\ra_{UHP} =0 \, .
\ee
This gives 
\be \label{PKfinal}
\la \BB | K_{2} |\phi\ra = 0
\ee
since $\la \BB |\chi\ra= \la f_{\BB} \circ\chi(0)\ra$ 
for any state 
$|\chi\ra$. This, in turn,
is equivalent to \refb{PK}.

\subsection{Oscillator representation of the butterfly state} 

We can also represent the matter part of the regulated butterfly state in 
the 
oscillator representation using eq.\refb{soscill}, \refb{Vbeta}.
In this case, with $\beta \equiv t^2$,
\be
v(\xi) = -\frac{\xi^3}{2}  \, ,\qquad
f_{\beta} (\xi)= \frac{\xi}{\sqrt{1 + \beta \xi^2}} \,.
\ee
Equ. \refb{Vbeta} gives
\ben \label{Vbetabutt}
\frac{d}{d \beta}  V_{mn}(\beta)& =& (-1)^{m+n} \frac{ \sqrt{mn}}{2}
\oint_0{dw\over 2\pi i}
\oint_0{dz\over 2\pi i}
\,{1\over z^{m+1} w^{n+1}}  \frac{ f_\beta(z)^3
- f_\beta(w)^3}{f_\beta(z)-f_\beta(w)}\\
&=& (-1)^{m+n} \frac{ \sqrt{mn} } {2}
\oint_0{dw\over 2\pi i}
\frac{f_\beta(w)}{w^{m+1}}
\oint_0{dz\over 2\pi i}  \frac{f_\beta(z)}{z^{m+1}}
= (-1)^{m+n} \frac{ \sqrt{mn}}{2}\,  x_m x_n \, , \nonumber
\een
where
\ben
x_m = \oint_0{dw\over 2\pi i}  \frac{f_\beta(w)}{w^{m+1}}
&=&(-\beta)^{\frac{m-1}{2}}
\frac{\Gamma[\frac{m}{2}] } {\sqrt{\pi}
\Gamma[ \frac{m+1}{2} ]}\quad {\rm for } \; m \;{\rm odd}\,,\\
&=& 0  \quad \qquad \qquad \qquad \qquad {\rm for }\;
m\; {\rm even}\,. \nonumber
\een
Integrating \refb{Vbetabutt} with the initial condition $V(\beta=0)=0$, we
find the Neumann coefficients of the regulated butterfly $(\beta \to 
t^2)$:
\ben \label{Vbutt}
V_{mn} (t) &=  &-(-1)^{\frac{m+n}{2}} \frac{\sqrt{mn}} {m+n}
\frac{\Gamma[\frac{m}{2}]
\Gamma[\frac{n}{2} ] } {\pi \Gamma[ \frac{m+1}{2} ]
\Gamma[\frac{n+1}{2}]}\, t^{m+n}\,,
\quad {\rm for } \; m  \; {\rm and}
\;n \;{\rm odd}\,,\\
&=& 0\,, \qquad \qquad \qquad \qquad \qquad \qquad \qquad \quad \quad 
\;\;\quad
{\rm for } \; m  \; {\rm or}
\;n \;{\rm even}\,. \nonumber
\een

\sectiono{The Nothing State} \label{snothing}

\begin{figure}[!ht]
\leavevmode
\begin{center}
\epsfysize=7cm
\epsfbox{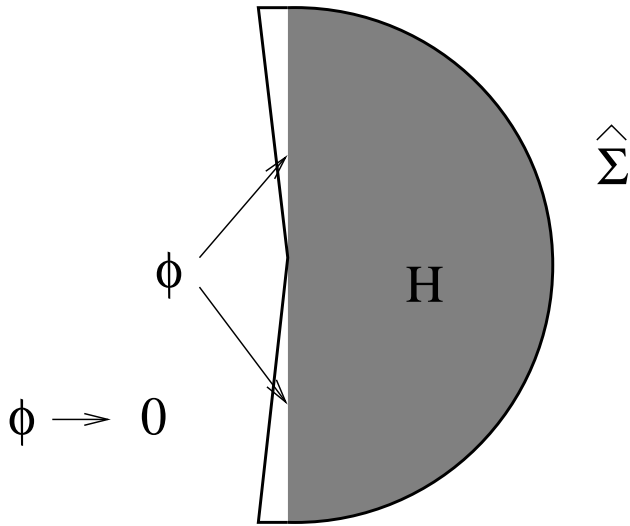}
\end{center}
\caption{The geometry of the disk $\wh\Sigma$ for the nothing state. 
Since the local coordinate patch fills the whole disk, the region $\wh 
\RR$, which represents the full disk $\wh\Sigma$ minus the local 
coordinate patch, collapses to nothing.} \label{fnothing} \end{figure}

The nothing state is defined by the relation:
\be \label{enoth1}
\la \NN|\phi\ra = \la f_\NN\circ \phi(0)\ra_{UHP}\,,
\ee
with
\be \label{enoth2}
f_\NN(\xi) = {\xi\over \xi^2 + 1}\, .
\ee
Under the map $\wh w(\xi)=h(\xi)$ with $h(\xi)$ defined as in 
eq.\refb{x2}, 
the upper half $z=f_\NN(\xi)$ plane gets mapped to the vertical half-disk 
$\wh\Sigma$ as shown in Fig.\ref{fnothing}. Clearly the boundary along the 
vertical line passes through the string mid-point which is at the origin 
of the $\wh w$-plane, and hence this 
state satisfies the usual criterion of being a projector of the 
$*$-algebra.

Various properties of the nothing state can be derived along the same 
lines 
as those of the butterfly. Here we summarize the main results:
\begin{itemize}
\item The nothing state factorizes into a product of the nothing state for 
the left half-string and the nothing state for the right half-string.
This follows from the results of section \ref{s3a.1p}, 
-- since in this case $\wh \RR$ associated with the original projector 
collapses, $\wh\RR_{top}$ also collapses and hence from \refb{toptohalf} 
it follows 
that $\wh\RR_{1/2}$ also collapses. Thus $\wh\Sigma_{1/2}$ is identical to 
$\wh\Sigma$. This proves that the half-string state is the same as the 
original state.

\item 
The map $f_{\NN} (\xi)$ defining the nothing state is related to the map
$f_{\II}(\xi)$ defining the identity string field by 
\be
f_{\NN}(\xi) = -i f_{\II}(i \xi) \,.
\ee
It follows from \refb{Uf}--\refb{vdef} that the operator expressions
of the identity and of the nothing state are related by
the formal replacement $L_{-2n} \leftrightarrow (-)^n L_{-2n}$.
The identity 
admits an elegant operator expression \cite{0105024} as an infinite 
product of exponentials of Virasoro generators. Changing
the sign of $L_{-2}$ in (3.3) of \cite{0105024}, we immediately have
\ben
|\NN \ra & = & \left( \prod\limits_{n=2}^{\infty} 
        \exp\left\{- \frac{2}{2^n} L_{-2^n}\right\} \right)
  e^{-L_{-2}} |0 \ra \nonumber \\
&=& 
\ldots
\exp(-\frac{2}{2^3} L_{-2^3}) \exp(-\frac{2}{2^2} L_{-2^2})
  \exp(-L_{-2}) |0 \ra  \,.
\een

\item 
$V^f_{mn}$ computed using \refb{evmn}, \refb{enoth2} turns out to be equal 
to $\delta_{mn}$. Thus
the oscillator representation of the matter part of the nothing 
state is given by:
\be \label{enoth4}
|\NN\ra_m = \exp \left(  -\frac{1}{2} \sum_{m,n=1}^\infty a^\dagger_n
a^\dagger_n \right) |0 \ra \,.
\ee

\item Computation of $A_f$ following 
eq.\refb{evfaf} 
gives 
$A_f=0$. Using eq.\refb{ewave} we then see that
the wave-functional of the nothing state, expressed as a functional 
of the coordinates $X_n$, is a constant independent of $X_n$.

\item  
The nothing state
is annihilated by all {\it even} reparametrizations of the cubic vertex,
\be \label{nothingK}
K_{2 n} | \NN \ra = 0 \qquad \forall n\, ,
\ee
where $K_{2n} = L_{2n} - L_{-2n}$. This is shown
with an argument similar to the one used for the
butterfly state, eqs.\refb{PK}-\refb{PKfinal}.  
Taking in this case the globally-defined vector fields 
\be
\tilde v_{2} (z) =-\frac{1}{z} + 4 z   \,, 
\quad \tilde v_4(z) =-\frac{1}{z^3}+\frac{6}{z} - 8 z \, ,
\ee
we find that
\be 
K_2 | \NN \ra =0 \, , \quad  K_4 |\NN \ra = 0 \,.
\ee
The commutation relations 
\be \label{Kalgebra}
[ K_m \, , K_n ] = (m-n) K_{m+n} - (-1)^n (m+n) K_{m-n}
\ee
then imply \refb{nothingK} for all $n$. Let us recall
that the identity string field is annihilated by
{\it all}, even and odd, vertex reparametrizations \cite{0006240},
so from this point of view the nothing state is the
most symmetric surface state apart from the identity.

\end{itemize}

\sectiono{The Generalized Butterfly States} \label{s2c} 

In this section we shall introduce a new class of surface states, $-$
called
generalized butterflies, $-$ each of which is a projector of the star
algebra. We shall first define these states, and then show that each of
these states satisfies the projector equation.
We shall also determine the half-string wave-functionals 
that the wave-functional of the generalized
butterfly state factorizes into.

\subsection{Definition of general butterflies} \label{s2.1n}

Let us begin by defining the generalized butterfly state
$|\BB_\alpha\rangle$. We generalize eq.\refb{defbut} to
\be \label{e2}
z= {1\over \alpha} \sin(\alpha \tan^{-1}\xi)\equiv 
f_\alpha(\xi)\, .  
\ee
As a result eq.\refb{moddef} is generalized to
\be \label{e3}
z
= {1\over \alpha} \sin(\alpha \wh z)\, .
\ee
Comparing eqs.\refb{moddef} and \refb{e3} we see that the generalized
butterfly differs from the original butterfly by a rescaling of the $\wh
z$ coordinate by a factor $\alpha$. 
Having a look at figure \ref{ff2}(c), we see that the generalized
butterfly occupies the region 
$-{\pi\over 2\alpha} < \Re (\wh z) \leq {\pi\over 2\alpha}$
in the upper half $\wh z$ plane. 
We denote by $\CC_\alpha$ 
this region in the $\wh z$ coordinate system, or more precisely,
a convenient translate of it.

As can be easily seen from eq.\refb{e2}, the map $f_\alpha(\xi)$ is
singular at the string mid-point $\xi=i$. In particular the mid-point is
sent to $i\infty$ and hence touches the boundary of the upper half
$z$-plane.
Thus from the general analysis of section \ref{s3a} we expect these states
to be projectors of the $*$-algebra and have factorized wave-functionals.
Also note that we have:
\be \label{efa1}
f_{\alpha=1} = {\xi\over \sqrt{1+\xi^2} }\, .
\ee
Comparison with eq.\refb{defbut} shows that the state
$|\BB_{\alpha=1}\ra$ is identical to the butterfly state defined in the
previous section. On the other hand, we have:
\be \label{efa}
f_{\alpha=0} = \tan^{-1}\xi\, .
\ee
This shows that the state $|\BB_{\alpha=0}\ra$ is identical
to the
sliver. The family of surface states $|\BB_{\alpha}\ra$ gives a family of
projectors, interpolating
between the butterfly and the sliver.
Finally, note that for $\alpha=2$ we have the map
\be \label{efa2}
f_{\alpha=2} = {\xi \over 1 + \xi^2}\, .
\ee
For reasons to be explained shortly, we call this the `nothing' 
state.

\begin{figure}[!ht]
\leavevmode
\begin{center}
\epsfysize=7cm
\epsfbox{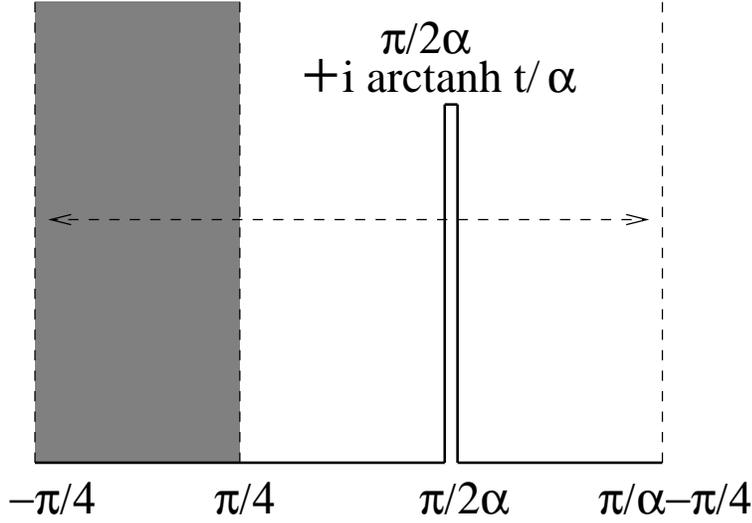}
\end{center}
\caption{The geometry of $\CC_{\alpha,t}$ in the complex $\wh z$ plane.
The shaded region denotes the
local coordinate patch, and the lines $\Re(\wh z)=-\pi/4$, $\Re(\wh
z)=\pi/\alpha-\pi/4$ are identified.} \label{f1}
\end{figure}
We can regularize the
singularity at the midpoint and define the
regularized butterfly
by generalizing \refb{defrbut} to
\be \label{e6}
z=f_{\alpha,t}(\xi) =
{1\over \alpha} { \tan(\alpha\tan^{-1}\xi) \over
\sqrt{1 + t^2 \tan^2 (\alpha\tan^{-1}\xi)} }
= {1\over \alpha} { \tan(\alpha\wh z)\over
\sqrt{1 + t^2 \tan^2 (\alpha \wh z) } }\, .
\ee
In the $\wh z$ plane we get
\be \label{e7}
\langle \BB_{\alpha,t}|\phi\rangle = \langle f^{(0)}\circ
\phi(0)\rangle_{
\CC_{\alpha,t}}\, ,
\ee
where $\CC_{\alpha,t}$ is the image of the upper half $z$ plane in the
$\wh z$ coordinate system and $f^{(0)}(\xi)=\tan^{-1}\xi$.
Comparison between \refb{defrbut} and \refb{e6} shows that the regularized
butterfly and the regularized generalized butterfly are related by a
rescaling of the $\wh z$ coordinate by $1/\alpha$. Thus
$\CC_{\alpha,t}$
can be obtained by a rescaling of Fig.\ref{ff3}(a) by $1/\alpha$, 
and moving the region to the left of the coordinate patch 
all the way to the right, as
shown in
Fig.\ref{f1}. Note that the local coordinate patch always occupies the 
same
region $|\Re(\wh z)|\le {\pi\over 4}$, $\Im(\wh z)\ge 0$,  
since $\wh z = \tan^{-1} \xi$.

\begin{figure}[!ht]
\leavevmode
\begin{center}
\epsfysize=6cm   
\epsfbox{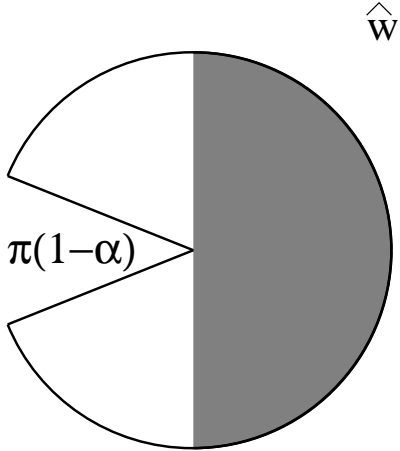}
\end{center}
\caption{The image of
$\CC_{\alpha}$ in the complex $\wh w=e^{2i\wh z}$ plane.
The shaded region denotes the
local coordinate patch.} \label{f2}
\end{figure}

\medskip  
As shown, $\CC_{\alpha,t}$ is a semi-infinite cylinder with circumference
$\pi/\alpha$, obtained by the restriction 
$\Im(\wh z)\ge 0$, 
$-\pi/4\le
\Re(\wh z) \le \pi/\alpha - \pi/4$, and the identification
$\Re(\wh z)=\Re(\wh z) + \pi/\alpha$ in the $\wh z$ plane, with a cut
along the
line $\Re(\wh z) = \pi/2\alpha$, 
extending all the way from the base $\wh
z=\pi/2\alpha$ to $\wh
z=\pi/2\alpha + i(\tanh^{-1}t)/\alpha$. As we move along the real $z$
axis, in the
$\wh z$ plane we go from $\wh z=-\pi/4$ to $\pi/2\alpha$ along the real
axis, then along the
cut to $\pi/2\alpha+i(\tanh^{-1}t)/\alpha$ and back to $\pi/2\alpha$, and
finally along
the real axis to $-\pi/4+\pi/\alpha$. The local coordinate patch,
corresponding to the unit half-disk in the upper half $\xi$ plane, is
mapped to the semi-infinite strip $\Im(\wh z)\ge 0$, 
$-\pi/4\le \Re(\wh z)\le
\pi/4$. The lines $\Re(\wh z)=\pi/4$ 
and $\Re(\wh z)=\pi/\alpha-\pi/4\equiv  
-\pi/4$ correspond to the images of the left and the right half of the
string respectively. As $t\to 1$ the cut goes all the way to
$\pi/2\alpha+i\infty$. The image of $\CC_{\alpha,t}$ in the complex $\wh
w=e^{2 i \wh z}$ plane in the $t\to 1$ limit has been shown in
Fig.\ref{f2}.

The tip of the cut at $\wh z = \pi/2\alpha+ i(\tanh^{-1}t)/\alpha$
corresponds to the branch point coming from the square root in the
denominator of \refb{e6}. According to our convention we choose the
positive sign of the square root to the left of this cut, $-$ this forces
us to choose the negative sign to the right of the cut. Thus in the
$t\to 1$ limit,  
the map from the $z$-plane to the $\wh z$ plane takes
the form:
\ben \label{eznew}
z = f_{\alpha,t=1}(\xi)  
&=& {1\over \alpha} \sin(\alpha\wh z) \qquad
\hbox{for} \quad \Re(\wh z) < \pi/2\alpha\, , \nonumber \\
&=& -{1\over \alpha} \sin(\alpha\wh z) \qquad
\hbox{for} \quad \Re(\wh z) > \pi/2\alpha\, .
\een
The difference from \refb{e3} for $\Re(\wh z)>\pi/2\alpha$ arises
because we redefined asymmetrically the fundamental domain in
the $\wh z$ plane.  
Note that for $\alpha=2$ the region of $\CC_{\alpha,t}$ outside the local
coordinate patch collapses to nothing.
For this  reason we call the
associated surface state the `nothing' state.

\begin{figure}[!ht]
\leavevmode
\begin{center}
\epsfysize=7cm
\epsfbox{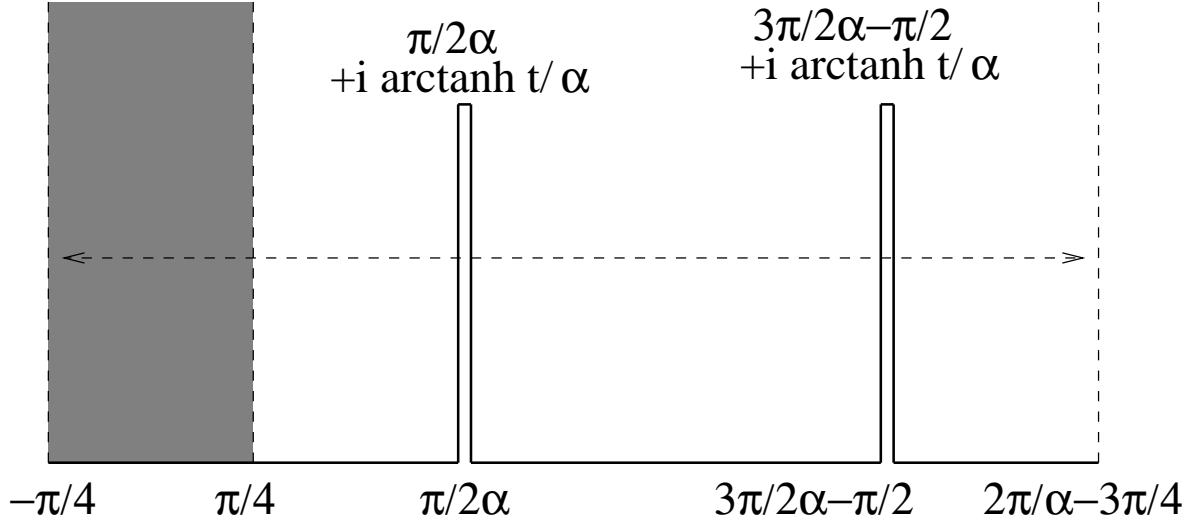}
\end{center}
\caption{The geometry of $\CC'_{\alpha,t}$ 
in the complex $\wh z$ plane.
The shaded region denotes the
local coordinate patch, and the lines $\Re(\wh z)=-\pi/4$, $\Re(\wh
z)=2\pi/\alpha-3\pi/4$ are identified.} \label{f3}
\end{figure}

\subsection{Squaring the generalized butterfly} \label{s2.2n}

We now want to show that $|\BB_\alpha\rangle$ squares to itself
under $*$ product. For this
we shall first compute $|\BB_{\alpha,t}*\BB_{\alpha,t}\rangle$,
and then take
the $t\to 1$ limit.
Throughout this section we shall work in
the combined matter-ghost system with zero central charge so that we can
apply the gluing theorem without any additional factors coming from
conformal anomaly. Since the discussion proceeds in a manner closely
parallel to that in section \ref{sregstar}, we shall omit the details and
only point out the essential differences.

As in section \ref{sregstar}, we work in the $\wh z$ coordinates.
We begin with
two copies of $
\CC_{\alpha,t}$ (one for each $\BB_{\alpha,t}$), simply remove the
local coordinate patch from the second
$
\CC_{\alpha,t}$ and glue the image of the
right half string on the first $
\CC_{\alpha,t}$ with the image of the left half string on the second $
\CC_{\alpha,t}$. The result is a cylinder 
$\CC'_{\alpha,t}$ of  
circumference
\be \label{ecircum}
{2\pi\over\alpha}-{\pi\over 2}\, ,
\ee
defined as the region $\Im(\wh z)\ge 0$, $-\pi/4\le
\Re(\wh z)\le 2\pi/\alpha-3\pi/4$. It also has two cuts, one along the
line $\Re(\wh z)=\pi/2\alpha$, extending from $\wh z=\pi/2\alpha$ to $\wh
z = \pi/2\alpha+i(\tanh^{-1}t)/\alpha$, and the other along the line
$\Re(\wh z) = 3\pi/2\alpha - \pi/2$, extending from $\wh z = 3\pi/2\alpha
- \pi/2$  to $\wh z = 3\pi/2\alpha
- \pi/2+ i(\tanh^{-1}t)/\alpha$. The local coordinate patch on 
$\CC'_{\alpha,t}$ is   
the vertical strip bounded by the lines $\Re(\wh z)=\pm\pi/4$. This has
been shown in Fig.\ref{f3}.

Let
$z=F_{\alpha,t}(\wh z)$ describes the map of 
$\CC'_{\alpha,t}$ to UHP.    
In order to show that $\BB_\alpha$ squares to itself, we need to
show
that as $t\to 1$, the map
$F_{\alpha,t}(\wh z)$ approaches the map given in \refb{e3} in
the  vicinity of the origin.
Thus the task is now to determine the map $F_{\alpha,t}$ that maps the
$\wh z$ plane to the upper half plane labeled by the coordinate $z$.
It is defined implicitly through the differential equation analogous to
\refb{starmap}:
\be \label{startmapnew}
d\wh z = {1\over (1+ d^2\, z^2)} \,{ (1 - b^2 z^2) \over \sqrt{
(1- a^2 z^2) (1-c^2 z^2) }}\, dz  \equiv 
H(z) dz\,.
\ee
The analog of eqs.\refb{con1}-\refb{con4} are:
\ben
\int_0^{1/a} H(z) dz  &=& {\pi\over 2\alpha}\,, \label{con1new}\\
\int_{1/a}^{1/b} H(z) dz  &=& ih/\alpha \,,\label{con2new}\\
\int_{1/b}^{1/c} H(z) dz  &=& -ih/\alpha \,,\label{con3new}\\
\int_{1\over c}^{\infty} H(z) dz  &=& {\pi\over 2\alpha} -
{\pi\over 4}
\,,\label{con4new}
\een
with $h$ as defined in eq.\refb{edefh}.

In the $t\to 1$ limit the height $h$ of the cylinder goes to $\infty$. We
shall now show that this can be achieved by taking
\be
\label{fonnew}
\{ b,c,d\} \ll a, \qquad c\ll\{b,d\} \,.
\ee
As in section \ref{sregstar} we define $z\to -1/z'$ and rewrite
\refb{startmapnew} as
\be\label{starmappnew}
d\wh z = {1\over (z'^2+ d^2)} \,{ (z'^2 - b^2 ) \over \sqrt{
(z'^2- a^2 ) (z'^2 -c^2 ) }}\, dz'  = G(z') dz'\,.
\ee
In the limit $\{b,c,d\}\ll a$,
\refb{con1new} gives:
\be
\int_a^\infty G(z') dz'  \simeq \int_a^\infty {dz'\over z'}
{1\over \sqrt{z'^2 - a^2 } } =  {\pi\over 2\alpha} \,.
\ee
This requires
\be\label{firstanew}
a \simeq \alpha \,.
\ee
Proceeding as in the case of section \ref{sregstar}, we can now show that
all the other conditions \refb{con2new}-\refb{con4new}, and the
requirement of
large $h$ can be satisfied by taking:
\be \label{efinalcondition}
c\ll b, d, \qquad d\sim b\, .
\ee
Using eqs.\refb{fonnew} and \refb{firstanew}, we see that
eq.\refb{startmapnew} now takes the form:
\be \label{starmapdonenew}
d\wh z =  \,{ 1  \over \sqrt{1-  \alpha^2 z^2 }}\, dz \,,
\ee
which gives:
\be \label{ezwhz}
z = {1\over \alpha} \sin(\alpha\wh z)\, .
\ee
This is precisely the map for the generalized butterfly. This establishes
that the generalized butterfly squares to itself under $*$-product.

\subsection{Wave-functionals for generalized butterfly states}  
\label{s3.2}

In this subsection we shall apply the general method described in section 
\ref{s3.1} to
compute the wave-functional of the generalized butterfly state.
In this process, we shall show explicitly that the wave-functional
factorizes into a
product of a functional of the left-half of the string and a functional of
the right-half of the string.
The wave-functional
of the butterfly state is expressed as
\be \label{ewb}
\langle \BB_\alpha|X\rangle = \NN_{\BB_\alpha} \exp \left( -\frac{1}{2}
\int_0^\pi \int_{0}^\pi  d\sigma d \sigma'
X(\sigma) A_{\BB_\alpha}(\sigma, \sigma') X(\sigma') \right) \, .
\ee
\begin{figure}[!ht]
\leavevmode
\begin{center}
\epsfysize=7cm
\epsfbox{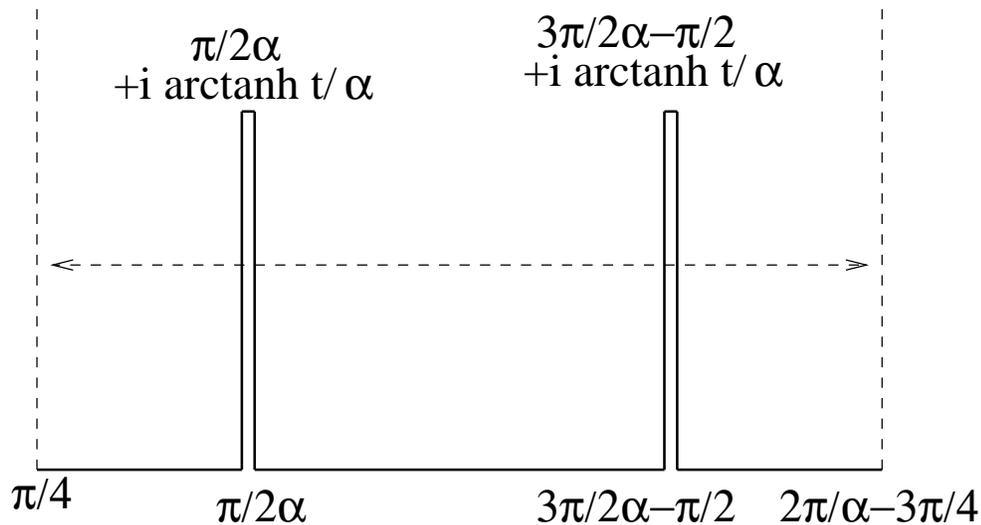}
\end{center}
\caption{The geometry involved in the computation of the inner product of
two generalized butterflies in the $\wh z$-plane.
} \label{f10}
\end{figure}
As seen from eqs.\refb{Bdef} and \refb{Bgauss},  
computation of $A_{\BB_\alpha}(\sigma_1, \sigma_2)$ defined in
eq.\refb{ewb} requires computing correlation functions of the form
$\langle \BB_{\alpha,t}|\cdots|\BB_{\alpha,t}\rangle$, and then taking the
limit $t\to 1$. To do
this computation one simply removes the local coordinate patch  from each
$\CC_{\alpha,t}$ and glues the
left half of one with the right half of the other and vice versa.
The
result is a semiinfinite cylinder $\CC''_{\alpha,t}$  
of circumference $(2\pi/\alpha -
\pi)$, corresponding to the region $\Im(\wh z)\ge 0$, 
$\pi/4\le \Re(\wh 
z)\le
\pi/4+2\pi/\alpha -\pi$, with the
identification $\wh z \equiv \wh z + (2\pi/\alpha -
\pi)$, and with two vertical cuts, one going from $\wh
z=\pi/2\alpha$ to $\wh
z = \pi/2\alpha+i(\tanh^{-1}t)/\alpha$, and the other going from $\wh z
= 3\pi/2\alpha
- \pi/2$  to $\wh z = 3\pi/2\alpha
- \pi/2+ i(\tanh^{-1}t)/\alpha$ respectively.
The two halves of the string along which we have glued the two
copies of $\CC_{\alpha,t}$ to produce the cylinder 
$\CC''_{\alpha,t}$ lie along the  
lines $\Re(\wh z)=\pi/4$ and $\pi/\alpha-\pi/4$ respectively. This has
been shown in Fig.\ref{f10}. 
Thus we have, using \refb{Bdef} and \refb{Bgauss}:  
\be \label{emnp}
{1\over 2} \p_{\sigma_1}
\p_{\sigma_2} A_{\BB_\alpha}^{-1}(\sigma_1,\sigma_2)
= \p_{\sigma_1} \p_{\sigma_2} \la
X(\wh z_1)
X(\wh z_2)\ra_{\CC''_{\alpha,t}}\, .  
\ee
This correlation function can
be calculated by finding the conformal transformation that maps the
cylinder $\CC''_{\alpha,t}$ 
to the upper half plane, and re-expressing 
\refb{emnp} as
a correlation function on the upper half plane.
This gives
\ben \label{e19}
{1\over 2} \p_{\sigma_1}
\p_{\sigma_2} A_{\BB_\alpha}^{-1}(\sigma_1,\sigma_2)
&=& \p_{\sigma_1} \p_{\sigma_2} \la
X(z_1) X(z_2) \ra_{UHP} \nonumber \\
&=& -{1\over 2} \p_{\sigma_1} \p_{\sigma_2} \{ \ln|z_1-z_2|^2 +
\ln|z_1-\bar z_2|^2\}\, ,
\een
with $z_i$'s computed in terms of $\sigma_i$'s with the help of the map
that relates $\wh z=\tan^{-1}\xi = \tan^{-1}(e^{i\sigma})$ to the upper
half plane coordinate $z$.
We could proceed as in section \ref{s2.2n} 
to construct the map from the
$\wh z$ plane to the $z$-plane, but in
this case we can write
down a closed form expression for this map.
The function that maps $\CC''_{\alpha,t}$ in the $\wh z$  
coordinate to the upper half plane labeled by $z$ is given by:
\be \label{e8}
z = g_{\alpha,t}(\xi) = g_{\alpha,t}(\tan\wh z)
\ee
with
\be \label{e9}
g_{\alpha,t}(\xi) = \sqrt{ \tan^2 (\beta\tan^{-1}\xi+\gamma) + u^2 \over
1 + u^2 \tan^2 (\beta\tan^{-1}\xi+\gamma) }
= \sqrt{ \tan^2 (\beta \wh z+\gamma)
+ u^2 \over 1 + u^2 \tan^2 (\beta\wh z+\gamma) }
\ee
where
\be \label{e10}  
{1\over \beta}={2\over \alpha} -1, \qquad \gamma = {\pi\over 2} (1-{\beta\over
\alpha}), \qquad {1\over \beta}\tanh^{-1}u = {1\over\alpha}\tanh^{-1}t\, .
\ee
To see that this maps $\CC''_{\alpha,t}$  
to the upper half plane,
we can start from $\wh z=\pi/4$ and follow the boundary of 
$\CC''_{\alpha,t}$  
to see that it maps to the real line in the $z$ plane. As we
start from
$\wh z=\pi/4$ and travel along the real axis
to $\wh
z =
\pi/2\alpha$, $z$ travels along the real line from 1 to 
$1/u$ 
As $\wh z$
goes from the point $\wh
z =
\pi/2\alpha$ towards $\pi/2\alpha + i (\tanh^{-1}t) / \alpha$,
$z$
goes from $1/u$ to $\infty$. Then as $\wh z$ returns back to $
\pi/2\alpha$ along the same line, $z$ goes from $-\infty$ to $-1/u$, and
as $\wh z$
travels along the real axis to $\wh z= (3\pi/2\alpha
-\pi/2)$, $z$ goes from
$-1/u$ to $-u$, passing through $-1$ at $\wh z = (\pi/\alpha -
\pi/4)$. As $\wh z$ travels along the vertical line from
$\wh z =
(3\pi/2\alpha -\pi/2)$ to $\wh z=(3\pi/2\alpha
-\pi/2)+i(\tanh^{-1}t)/\alpha$, $z$
goes from $-u$ to 0, and
as we return back to the point $\wh z =
3\pi/2\alpha -\pi/2$ along the same line, $z$ goes from 0 to
$u$. Finally, as we
move from $\wh z =
(3\pi/2\alpha -\pi/2)$ to $\wh z = (\pi/4 
+2\pi/\alpha -
\pi)$, $z$ goes from $u$ to 1. Since in the $\wh z$ plane we identify
the lines $\Re(\wh z)=(\pi/4+2\pi/\alpha -
\pi)$ with $\Re(\wh z) = \pi/4$, the contour closes.

{}From the map of
the boundary to the real $z$ axis described above we see that the ends of
the half strings, $\wh z=\pi/4$ and
$\wh z = (\pi/\alpha - \pi/4)$
maps to $1$ and $-1$ respectively. From eqs.\refb{e8}, \refb{e9},
\refb{e10}
we see that the half string $\Re(\wh z)=\pi/4$ gets mapped to
\be \label{e12}
z=\sqrt{(1+iS)^2 + u^2 (1-iS)^2 \over (1-iS)^2 + u^2 (1+iS)^2}\,,
\ee
where
\be \label{e13}
S =\tanh(\beta \Im(\wh z))\, .
\ee
On the other hand the half-string 
$\Re(\wh z) = (\pi/\alpha - \pi/4)$
gets mapped to
\be \label{e14}
z=-\sqrt{(1-iS)^2 + u^2 (1+iS)^2 \over (1+iS)^2 + u^2 (1-iS)^2} \,.
\ee
{}From eq.\refb{e10} we see that as 
 $t\to 1$ we have 
 $u\to 1$. Eqs.\refb{e12} and
\refb{e14} then show 
that
as $u\to 1$, the left and right half-strings are mapped in such a way
that all points of the left one, and all points on the right one,
except for the one associated to the full string midpoint, approach
the points $1$ and $-1$ respectively. 
The half strings remain infinite in the $z$ plane 
but are being reparametrized so
that all of their inner points are approaching either $1$ or $-1$. 
Eq.\refb{e19} now gives
\be \label{e16}
{1\over 2} \p_{\sigma_1}
\p_{\sigma_2} A_{\BB_\alpha}^{-1}(\sigma_1,\sigma_2)
=0\, \qquad \hbox{if} \quad 0\le\sigma_1<\pi/2, \quad
\pi/2<\sigma_2\le\pi\, .
\ee
Indeed, since the half-string points $\sigma_1$ and $\sigma_2$ 
are mapped to fixed points in 
the limit $t\to 1$, the  derivatives $dz_i/d\sigma_i$ go to zero,
and with $|z_1-z_2|$ finite, the evaluation of the right hand side of
Eq.\refb{e19} gives zero.
This is consistent with the wave-functional factorizing into
a product of a functional of the left half-string and a functional of the
right half-string. Indeed, since the interior of the right half-string 
is going into a point 
and the interior of the left half-string is going 
into another point in the $z$-plane, we have an explicit verification of
the factorization relation \refb{y2}.

In order to determine which particular state of the half-string appears in
the product, we need to evaluate the right hand side of \refb{e19} when
$\sigma_1$ and $\sigma_2$ lie on the same half of the string. For
definiteness we shall take
$0\le \sigma_1,\sigma_2<\pi/2$. The coordinate $\sigma$ of the half-string
is related to
$\wh z$ through $\wh z = \tan^{-1}(e^{i\sigma})$.
On the left half string, we can rewrite this as:
\be \label{e18}
\wh z = \pi/4 + {i\over 2} \ln\bigg({1 + \sin \sigma\over
\cos\sigma}\bigg)\, .
\ee
Using eqs.\refb{e13}, \refb{e18} we get:
\be \label{e22}
S(\sigma) = {
(1+\sin\sigma)^\beta
- \cos^\beta\sigma
\over
(1+\sin\sigma)^\beta +
\cos^\beta\sigma
}\, .
\ee
We can now use eqs.\refb{e19}, \refb{e14} and \refb{e22} to compute
$\p_{\sigma_1}
\p_{\sigma_2} A_{\BB_\alpha}^{-1}(\sigma_1,\sigma_2)$. Note from
eqs.\refb{e14}, \refb{e22} that
as $t\to 1$, the points $z_1$ and
$z_2$ come close together and $dz_i/ d\sigma_i$ also vanishes. Thus we
need to do this computation by keeping $t$ slightly away from 1 and then
take the limit $t\to 1$. To first order in $(1-t)$,
\be \label{e20}
z = 1 + i (1 - u^2) {S\over 1-S^2}\, .
\ee
Substituting this into \refb{e19}
we get
\be \label{e21}
\p_{\sigma_1}
\p_{\sigma_2} A^{-1}_{\BB_\alpha}(\sigma_1, \sigma_2) = \p_{\sigma_1}
\p_{\sigma_2}\bigg[-\ln
\bigg|{S(\sigma_1)\over
1-S(\sigma_1)^2}
- {S(\sigma_2)\over
1-S(\sigma_2)^2}\bigg|^2 - \ln \bigg|{S(\sigma_1)\over
1-S(\sigma_1)^2}
+ {S(\sigma_2)\over
1-S(\sigma_2)^2}\bigg|^2\bigg]\, .
\ee
This, in turn, gives us the wave-functional of the generalized butterfly
state through eq.\refb{ewb}.

As a special example we can consider the case of the butterfly state
$\alpha=1$. In this case we have $\beta=1$ and hence 
\be \label{esp1}
S(\sigma) = {(1 + \sin\sigma) - \cos\sigma
\over (1 +
\sin\sigma) + \cos\sigma }\, .
\ee
Substituting this into eq.\refb{e21} we get
\be \label{esp2}
\p_{\sigma_1}
\p_{\sigma_2} A^{-1}_{\BB_{\alpha=1}}(\sigma_1, \sigma_2) =
-\p_{\sigma_1}
\p_{\sigma_2}\Big[\ln|2\cos(2\sigma_1) -
2\cos(2\sigma_2)|\Big]\, , \qquad \hbox{for} \quad 0\le
\sigma_1,\sigma_2<\pi/2\, .
\ee
Comparing this with \refb{Bvac} we see that
after a rescaling $\sigma\to 2\sigma$
the half string
wave-functional coincides with the ground state wave-functional of the
string.
This is in accordance with the analysis of section \ref{sbutterwave}.

We could also try to derive the wave-functional of the nothing state by 
taking the $\alpha\to 2$ limit. From \refb{e10} we see that in this limit 
$\beta\to\infty$. Eq.\refb{e13} then gives,
\be \label{enno1}
S \simeq 1 - 2 e^{-2\beta \Im(\wh z)}\, .
\ee
Eqs.\refb{Bdef},
\refb{Bgauss}, \refb{e21} then gives:  
\ben \label{enno2}
\la\NN| \p_{\sigma_1} X(\sigma_1) \p_{\sigma_2} 
X(\sigma_2)|\NN\ra &=& \p_{\sigma_1}
\p_{\sigma_2} A^{-1}_{\BB_{\alpha=2}}(\sigma_1, \sigma_2) 
\nonumber \\
&\simeq& 
-\p_{\sigma_1}
\p_{\sigma_2} \Big[ \ln\Big|e^{2\beta \Im(\wh z_1)} - e^{2\beta \Im(\wh 
z_2)}\Big|^2 + \ln\Big|e^{2\beta \Im(\wh z_1)} + e^{2\beta \Im(\wh
z_2)}\Big|^2\Big] \nonumber \\
&\simeq& -\p_{\sigma_1}
\p_{\sigma_2} \ln \Big|1 - e^{-4\beta|\Im(\wh z_1)-\Im(\wh
z_2)|}\Big|^2 
\, .
\een
This clearly vanishes in the $\beta\to\infty$ limit for $\sigma_1\ne 
\sigma_2$. This is consistent with the fact that the wave-functional of 
the nothing state is a constant independent of $X(\sigma)$, since as we 
see from eq.\refb{Bgauss}, if $A_f$ vanishes, then the path integration 
over 
$X$ makes the expectation value of $\p_{\sigma_1} X(\sigma_1) 
\p_{\sigma_2}
X(\sigma_2)$ vanish for $\sigma_1\ne \sigma_2$ due to the $X(\sigma)\to 
-X(\sigma)$ symmetry at each point $\sigma$.

\sectiono{Other Projectors and Star Subalgebras} \label{s5}  

So far in this paper we have 
developed general properties
of split wave-functionals and projectors, and also 
discussed in detail
certain projectors, such as the butterfly and its generalizations,
and 
the nothing state.
Of these, the butterfly has the simplest representation as 
the exponential of a single Virasoro generator acting on the vacuum. 
In this section we exhibit other 
projectors whose
Virasoro representation 
is as simple as that of the butterfly.
We also discuss  subalgebras of surface states that generalize
the commutative wedge state subalgebra.  

\subsection{A class of projectors 
with simple Virasoro representation}\label{opr} 

The butterfly state, which is simply given as   
$\exp(-{1\over 2} L_{-2} ) |0\rangle$
(see Eq.\refb{Bexp}),
suggests the question whether there are other projectors
which 
can be written as an exponential involving a single Virasoro
operator.
To this end, we consider the vector fields
\be
 v_{(n)} (\xi)  = - {\beta\over n}\,\xi^{n+1} \, ,  
\ee
which generate the diffeomorphisms \cite{lpp}
\be z =   f_{(n)}(\xi) =  \exp \Bigl( v_{(n)} (\xi) {\partial\over 
\partial \xi}
\Bigr)\, \xi 
={\xi \over  (1+ \beta\, \xi^n)^{1/n}} \,.
\ee
The associated surface states are
\be
|B_n(\beta)\rangle =  \exp \Bigl( -{\beta\over n} 
(-1)^n L_{-n} \Bigr)   
|0\rangle \,.   
\ee
For even $n$ one can readily implement the projector condition
$f(\xi = \pm i) = \infty$ by a choice of the parameter $\beta$. Indeed,
this condition fixes
\be
\beta = - (-)^{n/2} \,, \qquad  n \, \, \hbox{even}\,.
\ee
We therefore obtain candidate projectors
\be \label{P2n}
|P_{2m}\rangle =  \exp \Bigl( (-1)^m{1\over 2m} L_{-2m} \Bigr)\,
|0\rangle\,.
\ee
The case $m=1$ is the canonical butterfly,
and the next projectors are
\be 
\exp\Bigl( {1\over 4} L_{-4} \Bigr) |0\rangle\,, \quad 
\exp \Bigl( -{1\over 6}
L_{-6} \Bigr) |0\rangle\,,\quad \exp \Bigl( {1\over 8}
L_{-8} \Bigr) |0\rangle
\,\,\, \cdots
\ee
and so on. These projectors obey the conservation law
\be \label{PKm}
K_{2m} |P_{2m}\rangle =0 \, ,
\ee
which is the obvious generalization of \refb{PK} and can
be proven in the same way considering
the global vector fields
\be
\tilde v_{2m}(z)=2 (-1)^{m+1} z - z^{-2m +1} \, .
\ee
It is interesting to note that $|P_4\rangle$ for example, is 
a state where the open string boundary condition chosen to define
the state does not hold at the string endpoint.  This is because
the map $f_4(\xi ) = \xi/(1-\xi^4)^{1/4}$ is singular at $\xi= \pm 1$.
The boundary of $\Sigma$ is discontinuous at the open string
endpoints, and the phenomenon discussed at the end of section \ref{s3a.1p}
occurs.

\subsection{Subalgebras of surface states annihilated by $K_n$}

The family of wedge states $|\WW_r\ra$ 
\cite{0006240, 0201095},
defined in the $z$ representation by the maps
\be \label{wedges}
z =\frac{r}{2}  \tan \Bigl(\frac{2}{r} \arctan(\xi) \Bigr) \, ,
\ee
obeys
$ K_1 |\WW_r \ra =0\, ,$
for all values of the
parameter $r$, $1 \leq r \leq \infty$. 
The wedge states
interpolate between the identity $|\WW_1 \ra \equiv |\II \ra$
and the sliver, $|\WW_\infty \ra \equiv |\Xi \ra$.
By analogy, it is natural to ask if there is
a family of states all annihilated
by $K_2$, interpolating between the identity and the butterfly,
and also containing the nothing state, see \refb{nothingK}. Indeed,
we have found such a family, defined
by the maps
\be \label{gmu}
z = g^{(2)}_\mu (\xi) = \frac{1}{\sqrt{4 \mu}} 
\left[1- \left(\frac{1 - \xi^2}{1+ \xi^2}
 \right)^{2 \mu}  \right]^{\frac{1}{2}} \,.
\ee
For $\mu = -1$ we recover the identity, for $\mu =1/2$ the canonical
butterfly and for $\mu = 1$ the nothing state. The condition
$g_\mu^{(2)} ( \pm i) = \infty$ is satisfied for $\mu \geq 0$, so 
according to our general arguments 
all the states with $\mu \geq 0$ are candidate projectors. 

\medskip

More generally, for any given integer $n$, we can look for  the family of
all 
surface states in the kernel of $K_n$. Since $K_n$ is a derivation,
each family will be closed under star-multiplication.
Let $z=g^{(n)}(\xi)$ be a map
that defines a surface state annihilated by $K_n$.
We require as usual that $g^{(n)}(\xi)$ has a regular
Taylor expansion in $\xi=0$ with $g^{(n)}(0) = 0$,
$\frac{dg^{(n)}(0)}{d\xi} = 1$. We can find
the general form of $g^{(n)}(\xi)$ by demanding
that the vector field 
\be
\tilde v_n(z) = \frac{dz}{d \xi} (\xi^{n+1} -
(-1)^n \xi^{-n+1})\,,
\ee be globally defined in the UHP 
-- this is
precisely the condition that the state is annihilated by $K_n$. 
Clearly
$\tilde v_n(z)$ must have
a pole of order $(n-1)$ at $z=0$. 
The most general form for such a globally defined vector field is
\be
\tilde v_n(z) = \frac{P_{n+1}(z)}{z^{n-1}} \, ,
\ee
where $P_{n+1}(z)$ is a  polynomial of order $(n+1)$ 
with non-vanishing
constant term. The order of $P_{n+1}$ is fixed by the
requirement that the vector field is regular
at infinity, $\lim_{z \to  \infty} z^{-2} 
\tilde v_n(z) = {\rm const}$.
Distinguishing between the cases of $n$ even or odd we
find that the two previous equations lead to 
the differential equations
\ben
\frac{1}{2n} \, d  \ln      \left( \frac{1 - \xi^{n}}{1+\xi^{n}}  \right) 
&=& \frac{z^{n-1} }{P_{n+1}(z)}\, dz
\qquad  {\rm for} \;n \;    {\rm even} \, ,
\\
\frac{1}{n} \, d  \arctan(\xi^n) & = & \frac{z^{n-1} }{P_{n+1}(z)}\, dz
\qquad {\rm for}\; n \; {\rm odd} \,.\nonumber
\een
Demanding that the    surface state is twist even requires  
that $z$ be an odd function of $\xi$, and this restricts the polynomial
$P_{n+1}$ to contain only  even powers of $z$. For $n=1$,
the most general twist    even solution is the 
family of maps   \refb{wedges} defining the wedge states; 
for $n=2$,
imposing again the twist  even condition, 
we find the one-parameter  family \refb{gmu}. 
Higher values of $n$ give  multi-parameter solutions.

\sectiono{Butterfly States Associated with General BCFT} \label{s4}

\begin{figure}[!ht]
\leavevmode
\begin{center}
\epsfysize=7cm
\epsfbox{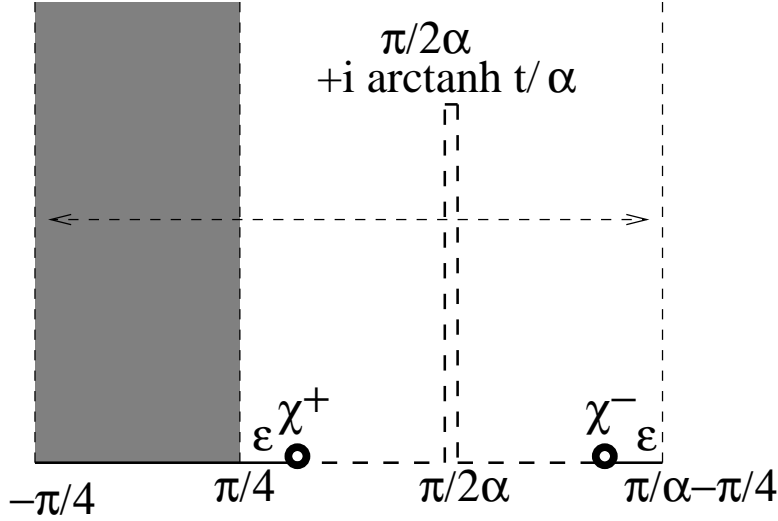}
\end{center}
\caption{The geometry involved in the computation of \refb{egen1}.
The two dots on the real line denote the insertion of $\chi^\pm$ at
distance $\epsilon$ away from the two edges. The thick dashed line on the
boundary represents $BCFT'$ boundary condition, whereas the thick
continuous line represents $BCFT$ boundary condition.} \label{f4}
\end{figure}

In Ref.\cite{0105168} we described the 
construction of sliver states associated
with a general boundary conformal field theory (BCFT). A very similar
construction can be carried out for generalized butterfly states 
associated with a general BCFT. For this we denote by $BCFT$ the reference
BCFT in
whose Hilbert space we wish to represent all the butterfly states, and by
$BCFT'$ some other BCFT. Let $\chi^\pm$ denote a pair of boundary
condition changing
operators of dimension $h$, such that an insertion of $\chi^+$ ($\chi^-$)
on the real axis separates $BCFT$ ($BCFT'$) boundary condition to the
left of the insertion
from $BCFT'$ ($BCFT$) boundary condition to the right of the insertion.
Furthermore, $\chi^\pm$ are required to
satisfy the operator product expansion:
\be \label{eope}
\chi^-(x) \chi^+(y) = (y-x)^{-2h} + \hbox{non-leading terms}\, .
\ee
Let us now define the state $|\BB_{\alpha,t}'\ra$ associated to a regulated
butterfly through the relation:
\be \label{egen1}
\la\BB_{\alpha,t}'| \phi\ra = (2\epsilon)^{2h} \Bigl\la f^{(0)}\circ\phi(0)
\chi^+({\pi\over 4 }+
\epsilon) \chi^-({\pi\over \alpha} - 
{\pi\over 4} - \epsilon)
\Bigr\ra_{\CC_{\alpha,t}}\, ,
\ee
where $\epsilon$ is any finite positive number,
$f^{(0)}(\xi)=\tan^{-1}\xi$, 
and $\CC_{\alpha,t}$ denotes the semi-infinite cylinder with
circumference $\pi/\alpha$ with a cut parametrized by $t$, as
shown in Fig.~\ref{f4}.
Let us define $|\BB_\alpha'\ra$ to be the state $|\BB_{\alpha,t=1}'\ra$
and $\CC_{\alpha}$ to be the cylinder $\CC_{\alpha,1}$.  
Using the conformal transformation
\be \label{egg1}
z = {1\over \alpha} \sin(\alpha\wh z)
\equiv g_\alpha(\wh z)\, ,
\ee
that maps $\CC_\alpha$ to the upper-half $z$-plane, 
we can reexpress $\la\BB_{\alpha}'|\phi\ra$ as
\be \label{egg2}
\la\BB_{\alpha}'| \phi\ra = (2\epsilon)^{2h} \Bigl\la
g_\alpha\circ f^{(0)}\circ\phi(0) \,
\,\,g_\alpha\circ \chi^+({\pi\over 4} +
\epsilon) \,\,\, g_\alpha\circ \chi^-(-{\pi\over 4} -
\epsilon)\Bigr\ra_{UHP}\, .
\ee
In the last step we have used the periodicity in the $\wh z$ plane to
replace $\chi^-(\pi/\alpha - 
\pi/4-\epsilon)$ by $\chi^-(-\pi/4-\epsilon)$.

\begin{figure}[!ht]
\leavevmode
\begin{center}
\epsfysize=6cm
\epsfbox{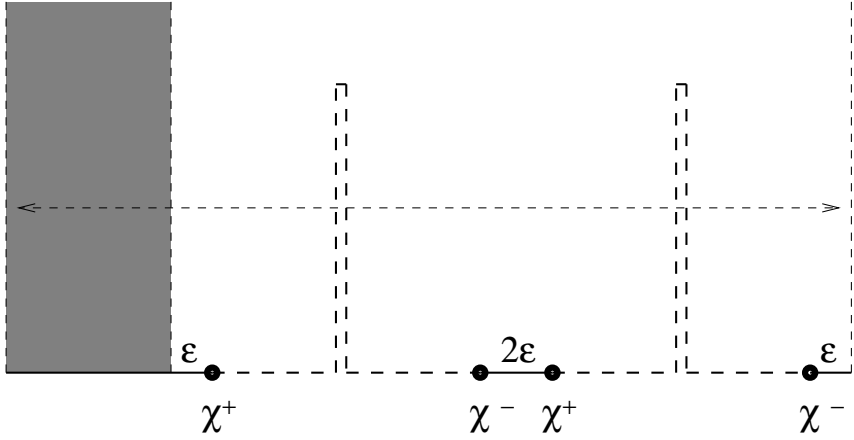}
\end{center}
\caption{The geometry involved in the computation of
$|\BB'_{\alpha,t}*\BB'_{\alpha,t}\ra$. The boundary components labeled by
thick continuous line represent $BCFT$ boundary conditions and the
boundary
components labeled by thick broken lines label 
$BCFT'$ boundary  
conditions. The coordinate labels of various points are identical to those
in Fig.\ref{f3}.} \label{f5}
\end{figure}

The computation of $|\BB'_{\alpha,t}*\BB'_{\alpha,t}\ra$ is
straightforward
using the gluing rules in the $\wh z$ coordinate system. The result is:
\be \label{egen2}
\la \BB_{\alpha,t}'*\BB_{\alpha,t}'|\phi\ra =
(2\epsilon)^{4h} \Bigl\la
f\circ\phi(0)
\chi^+({\pi\over 4}+\epsilon) 
\chi^-({\pi\over \alpha} -{\pi\over 4}-\epsilon)
\chi^+({\pi\over \alpha}-{\pi\over 4}+\epsilon) \chi^-({2\pi\over \alpha}-
{3\pi\over 4}-
\epsilon)\Bigr\ra_{\CC'_{\alpha,t}}\, .   
\ee
The geometry of $\CC'_{\alpha,t}$ 
has been shown in Fig.\ref{f5}. This 
correlation
function can
be evaluated by mapping it to the UHP via the map given in
eq.\refb{startmapnew}.
As shown in section \ref{s2.2n}, as $t\to 1$ this approaches
eq.\refb{ezwhz},
with part of the $\CC'_{\alpha,t}$ 
plane in the middle collapsing to a 
point in the
$\wh z$ plane. In this case this part includes the insertion
$\chi^-(\pi/\alpha-\pi/4-\epsilon)
\chi^+(\pi/\alpha-\pi/4+\epsilon)$, and hence this can be replaced by the
leading term in the operator product expansion which is
$(2\epsilon)^{-2h}$. Thus in the $t\to 1$ limit, 
we can rewrite
\refb{egen2} as,
\be \label{egen3}
\la \BB'_{\alpha}*\BB_{\alpha}'|\phi\ra = 
(2\epsilon)^{2h} \Bigl\la 
f\circ\phi(0)
\chi^+({\pi\over 4} +\epsilon) \chi^-(-{\pi\over 4} - 
\epsilon)\Bigr\ra_{\CC'_{\alpha,1}}\, ,  
\ee
where 
we have used the periodicity in the $\wh z$ plane to
replace $\chi^-(2\pi/\alpha-3\pi/4-
\epsilon)$ by $\chi^-(-\pi/4-\epsilon)$. 
Using the result of section \ref{s2.2n} that the map
$g_\alpha(\wh z)$ defined in eq.\refb{egg1}, that maps
$\CC_\alpha$ to the upper half plane, also maps 
$\CC'_{\alpha,t}$ to the 
upper half
plane for $t=1$, we have:
\be \label{egen6}
\la \BB'_{\alpha}*\BB_{\alpha}'|\phi\ra = (2\epsilon)^{2h} \Bigl\la
g_\alpha\circ f\circ\phi(0)
g_\alpha\circ \chi^+({\pi\over 4}+\epsilon)
g_\alpha\circ \chi^-(-{\pi\over 4}-\epsilon)\Bigr\ra_{UHP}\, .
\ee
Comparing eqs.\refb{egg2} and \refb{egen6} we see that
\be \label{egen7}
\la \BB'_\alpha| = \la \BB'_\alpha*\BB'_\alpha|\, ,
\ee
as we wanted to show.  We end this section by noting that 
this construction can be easily 
generalized to construct 
any of the projectors discussed in section \ref{new5} 
associated with a general BCFT.

\sectiono{Concluding Remarks}\label{concl} 

In this paper we have given a rather general discussion
of projectors of the open string $*$-algebra 
and split wave-functionals. It was found that 
in addition to the sliver, infinitely many projectors
exist  which  have 
pure geometrical interpretation as
surface states.  We have also seen that such surface
states, in general, can be viewed as tensor products
of half-string surface states. This viewpoint makes it
clear that the half-string surface states 
are naturally defined with  the 
same open string boundary conditions as the full string
states. 
Moreover, all projectors are clearly recognized
as being invariant under
opposite constant translations of the half strings.
We have illustrated in detail our general considerations,
by  discussing explicitly several interesting projectors.

While we have indicated that the projectors considered here
are expected to be equivalent in that they define physically
equivalent solutions of vacuum string field theory, our focus
on specific projectors and their properties may have applications
in other contexts. For example, it seems clear that OSFT
solutions are not projectors. It follows, that if OSFT solutions
are eventually built in terms of deformations of projectors,
particular projectors could be of special use. As noted, level
expansion does seem to single out the butterfly as a special
projector. There may still be other surface state projectors 
deserving  particular attention which 
we have not uncovered. 

One question we have not attempted to investigate is how
the approach of the midpoint to the boundary controls
the behavior of the projector. While $f(\xi=i) = \infty$
seems necessary to have a projector, it may be of interest to 
understand the full significance 
the behavior of $f$ near $\xi=i$. We have already seen that
this behavior controls the boundary conditions 
satisfied by the half string states.  

In this paper we have only considered string fields which 
have a purely geometric interpretation as surface states 
associated to Riemann surfaces. All such string fields
belong to the Virasoro module on the vacuum. We have
uncovered the general geometric mechanism that gives
rise to rank-one surface state projectors.
It should be kept in mind that there 
are many projectors that lie outside the Virasoro
module of the vacuum and do not have a purely geometric
interpretation. For example there are squeezed
states built with flat space oscillators
that star-multiply to themselves but are not
surface states.

It has been recently recognized \cite{ halfn,  0202087} 
that the open string star product
can be interpreted as a continuous tensor product of mutually commuting
two-dimensional Moyal products. This algebraic approach
is likely to shed new light on projectors,
and it will be interesting to understand in detail the connection
with the geometric methods of the present paper.
A complete understanding of projectors could well
help eventually give a concrete description of the star-algebra.
In fact, given the central role of projectors in the study of
non-commutative field theory, it is natural to expect that
star-algebra projectors will have an important role in our future
understanding of string field theory.

\bigskip  
\noindent
{\bf Acknowledgements.}  
We would like to thank M.~Schnabl
for discussions on the matters presented here.
The work of L.R. was supported in part
by Princeton University
``Dicke Fellowship'' and by NSF grant 9802484.
The  research of A.S. was supported in part by a grant 
from the Eberly College
of Science of the Penn State University.
The work of  B.Z. was supported in part
by DOE contract \#DE-FC02-94ER40818.

\appendix
\sectiono{Numerical Computations Involving the Butterfly}
\label{snum}

In this appendix we shall present numerical results for computations
involving the butterfly at various levels of approximation. We approximate
the butterfly by truncating it to a given level $L$, and calculate various
star products keeping only terms up to level $L$. 
The results are given in Tables 1, 2 and 3.
The last column gives the expected answers, and the last but one column
gives the extrapolation of the numerical answers to infinite level using
a fit of the form $a + b/L + c/L^2$.

Table 1 contains numerical results for the square of the butterfly.
As we see from this table, as we include more and more terms
in the expression for the butterfly, the closer 
is its square to
the expected answer.

We can also test the property \refb{rone} that the butterfly is
a rank-one projector. We need to work in a unitary BCFT, so 
we choose to consider butterfly states $|\BB^{c=1}\ra$ built 
with Virasoro generators
of central charge one. Star multiplication
of surface states in this theory works as in the
$c=0$ case but with extra (infinite) overall factors arising
from the conformal anomaly. These infinite
factors are regulated by the level truncation procedure
and we ignore them, considering the results
up to their overall normalization.
In Table 2 and Table 3 we present
the numerical results for the normalized products $|\BB^{c=1}\ra
* |0\ra * |\BB^{c=1}\ra$ and $|\BB^{c=1}\ra
* L_{-2} |0\ra * |\BB^{c=1}\ra$. From
\refb{rone} we expect these products to
be proportional to the butterfly,
and indeed this is seen to hold more and more accurately
as the level is increased.

These numerical results confirm the formal arguments of this paper that the
butterfly is a rank-one projector of the $*$-algebra.

\begin{table}
\begin{center}\def\st{\vrule height 3ex width -2ex}
\begin{tabular}{|l|l|l|l|l|l|l|l|l|} \hline
$ $ & $L=2$ & $L=4$ & $L=6$ & $L=8$  & $L=10$ & $L=12 $ & $L
=\infty$ & $Exp$
\st\\[1ex]
\hline \hline 
 $|0\ra$  & 1.00000 & 1.00000 & 1.00000 & 1.00000 &
1.00000 & 1.00000 & 1.00000 & 1.00000
\st\\[1ex]\hline
$L_{-2}|0\ra$  & -0.43230 & -0.46035 & -0.47214 &
-0.47858 & -0.48263 & -0.48540 & -0.49981 & -0.50000
\st\\[1ex]\hline $L_{-4}|0\ra$ & 0 & -0.00351
& -0.00213 & -0.00146 & -0.00108 & -0.00084 & 0.00021 & 0.00000
\st\\[1ex]\hline $L_{-2}L_{-2}|0\ra$ & 0 & 0.10845
& 0.11309 & 0.11567 & 0.11732 & 0.11847 & 0.12307 & 0.12500
\st\\[1ex]\hline $L_{-6}|0\ra$ & 0 & 0 & 0.00168
& 0.00117 & 0.00087 & 0.00068 & 0.00005 & 0.00000
\st\\[1ex]\hline $L_{-4}L_{-2}|0\ra$ & 0 & 0 & 0.00040
& 0.00027 & 0.00021 & 0.00017 & 0.00003 & 0.00000
\st\\[1ex]\hline $L_{-3}L_{-3}|0\ra$ & 0 & 0 & -0.00034
& -0.00023 & -0.00016 & -0.00012 & 0.00001 & 0.00000
\st\\[1ex]\hline $(L_{-2})^3|0\ra$ & 0 & 0
& -0.01771 & -0.01843 & -0.01887 & -0.01918 & -0.02017 & -0.02083
\st\\[1ex]
\hline
\end{tabular}
\end{center}
\caption{ Numerical results for the coefficients of
$|\BB\ra * |\BB\ra$. } 
\end{table}

\begin{table}
\begin{center}\def\st{\vrule height 3ex width -2ex}
\begin{tabular}{|l|l|l|l|l|l|l|l|l|} \hline
$ $ & $L=2$ & $L=4$ & $L=6$ & $L=8$  & $L=10$ & $L=12 $ & $L
=\infty$ & $Exp$
\st\\[1ex]
\hline \hline 
 $|0\ra$  & 1.00000 & 1.00000 & 1.00000 & 1.00000 &
1.00000 & 1.00000 & 1.00000 & 1.00000
\st\\[1ex]\hline
$L_{-2}|0\ra$  & -0.39892 & -0.44581 & -0.46296 &
-0.47192 & -0.47743 & -0.48115 & -0.50002 & -0.50000
\st\\[1ex]\hline $L_{-4}|0\ra$ & 0 & 
-0.00617 & -0.00231 & -0.00136 & -0.00093 & -0.00069 & 0.00009 & 0.00000
\st\\[1ex]\hline $L_{-2}L_{-2}|0\ra$ & 0 & 0.09982 
& 0.10808 & 0.11217 & 0.11465 & 0.11633 & 0.12312 & 0.12500
\st\\[1ex]\hline $L_{-6}|0\ra$ & 0 & 0 & 0.00349 
& 0.00143 & 0.00088 & 0.00063 & -0.00002 & 0.00000
\st\\[1ex]\hline $L_{-4}L{-2}|0\ra$ & 0 & 0 
& 0.00117 & 0.00054 & 0.00033 & 0.00024 & -0.00005 & 0.00000
\st\\[1ex]\hline $L_{-3}L_{-3}|0\ra$ & 0 & 0 
& -0.00005 & -0.00011 & -0.00009 & -0.00007 & 0.00000 & 0.00000
\st\\[1ex]\hline $L_{-2}L_{-2}L_{-2}|0\ra$ 
& 0 & 0 & -0.01652 & -0.01761 & -0.01825 & -0.01867 & -0.02007 & -0.02083
\st\\[1ex]
\hline
\end{tabular}
\end{center}

\caption{ Numerical results for the coefficients of
$|\BB^{c=1}\ra
* |0\ra * |\BB^{c=1}\ra$. The coefficient of $|0\ra$ in the result
has been normalized to one.} 
\end{table}

\begin{table}
\begin{center}\def\st{\vrule height 3ex width -2ex}
\begin{tabular}{|l|l|l|l|l|l|l|l|l|} \hline
$ $ & $L=2$ & $L=4$ & $L=6$ & $L=8$  & $L=10$ & $L=12 $ & $L
=\infty$ & $Exp$
\st\\[1ex]
\hline \hline $|0\ra$  & 1.00000 & 1.00000 & 1.00000 & 1.00000 &
1.00000 & 1.00000 & 1.00000 & 1.00000
\st\\[1ex]\hline $L_{-2}|0\ra$ & -0.48377 & -0.49010 & -0.48462 & -0.48562 
& -0.48720 & -0.48862 & -0.48519 & -0.50000
\st\\[1ex]\hline $L_{-4}|0\ra$ & 0 & -0.05238 
& -0.00861 & -0.00461 & -0.00312 & -0.00233 & -0.00689 & 0.00000
\st\\[1ex]\hline $L_{-2}L_{-2}|0\ra$ & 0 & 
0.11387 & 0.11410 & 0.11565 & 0.11712 & 0.11823 & 0.12254 & 0.12500
\st\\[1ex]\hline $L_{-6}|0\ra$ & 0 & 0 & 0.02146 
& 0.00456 & 0.00247 & 0.00166 & 0.00258 & 0.00000
\st\\[1ex]\hline $L_{-4}L{-2}|0\ra$ & 0 & 0 
& 0.00288 & 0.00054 & 0.00043 & 0.00035 & 0.00086 & 0.00000
\st\\[1ex]\hline $L_{-3}L_{-3}|0\ra$ & 0 & 0 
& -0.00083 & -0.00093 & -0.00055 & -0.00038 & 0.00059 & 0.00000
\st\\[1ex]\hline $L_{-2}L_{-2}L_{-2}|0\ra$ & 0 & 0 
& -0.01738 & -0.01795 & -0.01848 & -0.01886 & -0.02016 & -0.02083
\st\\[1ex]
\hline
\end{tabular}
\end{center}
\caption{ Numerical results for the coefficients of
$|\BB^{c=1}\ra
* L_{-2} |0\ra * |\BB^{c=1}\ra$. The coefficient of $|0\ra$ in the result
has been normalized to one.} 
\end{table}

\end{document}